\documentclass[%
reprint,
 amsmath,amssymb,
 aps,
 showkeys,
]{revtex4-2}

\usepackage{graphicx}
\usepackage{dcolumn}
\usepackage{bm}

\usepackage[dvipsnames]{xcolor}

\usepackage{xcolor}
\usepackage{soul}
\definecolor{reddish}{HTML}{FBB4AE}
\definecolor{blueish}{HTML}{B3CDE3}
\definecolor{magentish}{HTML}{FF00AA}
\definecolor{greenish}{HTML}{a1d99b}

\graphicspath{{figures/}}

\usepackage{booktabs}

\begin{document}

\title{
    Combining sensors and surveys to study social contexts:\\
    Case of scientific conferences
    }

\author{Mathieu G\'enois}
\email{mathieu.genois@cpt.univ-mrs.fr}
\affiliation{Aix Marseille Univ, Université de Toulon, CNRS, CPT, Marseille, France}
\affiliation{GESIS -- Leibniz-Institut f\"ur Sozialwissenschaften, Köln, Germany}
\author{Maria Zens}%
\affiliation{GESIS -- Leibniz-Institut f\"ur Sozialwissenschaften, Köln, Germany}
\author{Marcos Oliveira}%
\affiliation{Computer Science, University of Exeter, Exeter, UK}
\author{Clemens M. Lechner}%
\affiliation{GESIS -- Leibniz-Institut f\"ur Sozialwissenschaften, Mannheim, Germany}
\author{Johann Schaible}%
\affiliation{TH-Köln -- University of Applied Sciences, Gummersbach, Germany}
\author{Markus Strohmaier}%
\affiliation{Business School, University of Mannheim, Germany}
\affiliation{GESIS -- Leibniz-Institut f\"ur Sozialwissenschaften, Köln, Germany}
\affiliation{Complexity Science Hub Vienna, Austria}

\date{\today}

\begin{abstract}
In this paper, we present a unique collection of four data sets to study social behaviour. The data were collected at four international scientific conferences, during which we measured face-to-face contacts along with additional information about individuals. Building on innovative methods developed in the last decade to study human social behaviour, interactions between participants were monitored using the SocioPatterns platform, which allows collecting face-to-face physical proximity events every 20 seconds in a well-defined social context. Through accompanying surveys, we gathered extensive information about the participants, including sociodemographic characteristics, Big Five personality traits, DIAMONDS situational perceptions, measure of scientific attractiveness, motivations for attending the conferences, and perceptions of the crowd (e.g., in terms of gender distribution). Linking the sensor and survey data provides a rich window into social behaviour: At the individual level, the data sets allow personality scientists to investigate individual differences in social behaviour and pinpoint which individual characteristics (e.g., social roles, personality traits, situational perceptions) drive these individual differences. At the group level, the data also allow studying the mechanisms responsible for interacting patterns within a scientific crowd during a social, networking and idea-sharing event. The data are available for secondary analysis.
\end{abstract}


\maketitle

\section{Introduction}

The study of human behaviour now includes device-based quantitative methods enabling researchers to comprehend our behaviour better. Many of these novel methods have emerged with the expansion of electronic and online media, particularly mobile phones and the Internet. At the same time that such media increasingly shape our interpersonal behaviour, they provide us with the means to collect fine-grained data about human behaviour, which can be used for answering research questions.

For example, the ubiquity of mobile phones enables us to understand how individuals interact~\cite{calabrese:2014}, and researchers have used it as a proxy for individuals' geographical location to investigate spatial crowd dynamics~\cite{calabrese:2014,rojas:2016}. Likewise, the diffusion of GPS as an everyday tool was another step in the development of methods to probe human travel patterns~\cite{rout:2021,sila2016analysis}. Remarkably, the Internet and its multiple usages have introduced new tools that open another window on human behaviour (e.g., online social networks, instant messaging, web browsing). In all these tools, the common feature is that they generate \emph{digital traces}, data about the users' behaviour that can be automatically collected and stored.

Computational social scientists have rapidly noticed how they could use these data sources to study human behaviour, particularly data about online behaviour. Because of its vast availability, online data have been extensively used to investigate human behaviour, becoming the leading research topic for human interactions. However, such research efforts have the caveat that results obtained with online media may not necessarily be transposed onto their real-world counterparts~\cite{mellon:2017}. Crucially, new questions arise: for example, do electronic and verbal communications share common properties? Are social circles similar online and offline? How do online and offline behaviour translate and impact one another?

To tackle these and other critical questions, we need to be able to probe the real world in the same quantitative way as the online world. To that end, researchers have developed \emph{sensors}, either relying on existing infrastructure---usually smartphones~\cite{stopczynski:2014,vu:2010}---or designing their own~\citep{choudhury:2003,salathe:2010}. Such sensors detect physical proximity between participants, which constitutes a proxy for social contacts~\cite{Routledge2022,malik:2018}. This redefinition of behaviour measurement allows for collecting quantitative information about how individuals interact with each other in physical space.

We have thus a new tool to study human social behaviour \emph{in situ}, which gives new ways to look at the phenomena at play. By pairing sensor data with surveys, we can objectively measure, and quantify individual differences in, social behavior. Moreover, we can pinpoint individual and contextual characteristics that shape social behaviour and underline individual differences therein. We can thus contribute to a better understanding of the linkage between personality and social behaviour, a topic that has received considerable attention in personality science in recent years (e.g., ~\cite{back2021, breil2019}).

In this work, we focus on scientific conferences as an example of a social context where social interactions can be driven by several factors: social roles and social status, personality traits, situational perceptions, and motivations to cite but a few. Such data sets allow for a wide range of exploratory studies regarding the effect of each of these factors on contact behaviour, correlations between them, insights into crowd dynamics in the sociology of science, and general properties of contacts between individuals in different contexts. 

\section{Methods}

This section presents all details of the data collection procedure.

\subsection{Contact data}

\subsubsection{The SocioPatterns platform}

The first part of each study consists in recording interactions between participants. A \emph{social interaction} can include many different behaviours, such as conversation, physical contact, and eye contact. All are relevant for the analysis of ties within a crowd. In the present case, we focus on the more straightforward, broader definition of a \emph{contact} as a physical, face-to-face proximity event. Although physical proximity between individuals does not necessarily imply an interaction, previous work shows that this signal constitutes an excellent proxy, which enables the analysis of the structure of a social context~\citep{Routledge2022}.

We used the SocioPatterns platform \citep{Cattuto:PLOS2010} to collect contacts between participants, which has been largely used in the past decade to explore interaction patterns in social contexts~\cite{vanhems:2013,kiti:2016,EPJDS:2018,kontro:EducSci2020,ozella:2021,oliveira_group_2022}. This equipment consists of sensors attached to the participants' name tags and antennas covering the conference venue to collect contact data from the sensors. Each sensor carries an RFID chip and can detect other sensors in the vicinity within a $\sim$\,1.5\,m radius. Furthermore, as the human body blocks the emitted signal, detection only occurs when two individuals are face-to-face (i.e., in their respective front half-spheres). An event with such proximity and geometry defines a \emph{contact}. Contacts are recorded every 20 seconds and are limited to 40 simultaneous contacts for each individual within a 20-seconds time window. By design, contacts lasting at least 20 seconds have $\sim$100\,\% chance of being recorded. Shorter contacts may be recorded, with a probability decreasing with their duration.

\subsubsection{Setting up the contact tracking platform}

As sensors only have limited memory, antennas are necessary to collect the data from them continuously. Coverage of the conference venue is thus crucial to ensure that the maximum amount of contacts is collected. Antennas have a theoretical detection radius of $\sim$30\,m. Thus, we examined each conference venue floor plan to identify the suitable number of antennas needed. Because sensors and antenna communicate via radio waves, we performed tests \emph{in situ} to evaluate the impact of obstacles, in particular walls and windows. See Appendix~\ref{app:plans} for a detailed description of the coverage of each venue.

\subsubsection{Participation and sensor distribution}

Participation was offered to all attendees of the conferences upon registration (usually online before the event); attendees could opt out at their arrival at the event. Table~\ref{tab:stat} summarises the resulting participation rates.

\begin{figure}
    \centering
    \includegraphics[width=.5\columnwidth]{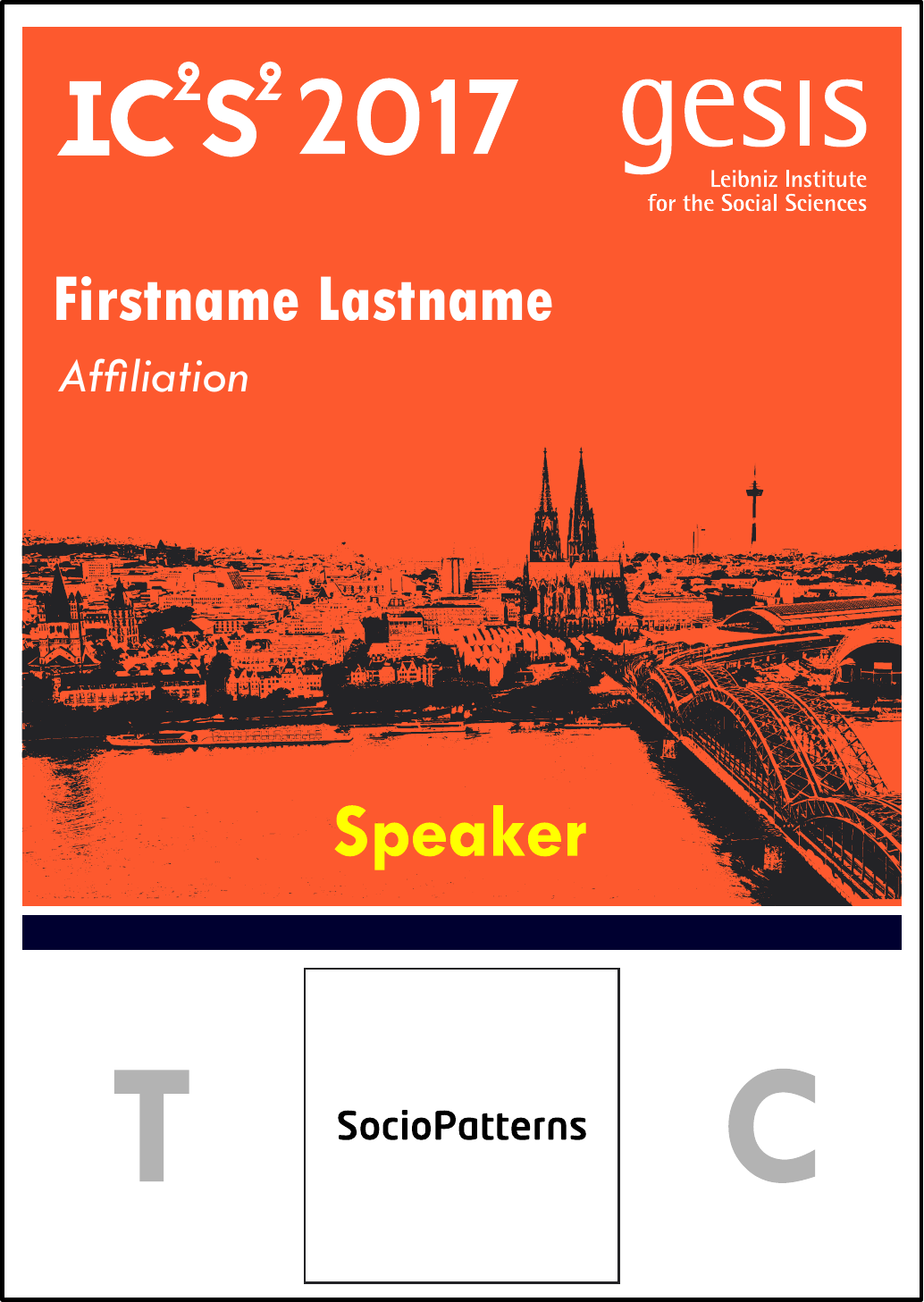}
    \caption{\textbf{Example of name tag.} The square frame at the bottom was reserved to receive the SocioPatterns sensor. The name tag is 105x148 mm, the sensor frame is 36x36 mm.}
    \label{fig:badge}
\end{figure}

To avoid manipulation by the participants, we preemptively installed sensors within the name tags used for the conferences (see Fig.~\ref{fig:badge}). Before the conference, we sent an e-mail to all participants informing them that a SocioPatterns study was taking place during the conference, attached with a form of consent with a complete description of the data collection (see Appendix~\ref{app:consent}). Upon registration at the conference, participants could choose to participate or refuse; a data collection team member was also available to answer questions. If they agreed, they were given a form of consent to sign. If they refused, the sensor was removed from the name tag. When leaving at the end of a conference day, participants kept their name tags with the sensor and brought them back the next day. We note that no contact detection occurs outside the conference venue. Upon leaving the conference permanently, the participant returned their name tag to the registration desk.

\subsubsection{Data cleaning}

The raw data gathered by the antennas first goes through a preprocessing phase, in which the contacts are \emph{aligned}. This process is necessary because neither sensors nor antennas include an internal clock. Their data thus have to be synchronised. Furthermore, the data are \emph{binned} into 20 seconds time windows.

\begin{figure}[h]
    \centering
    \includegraphics[width=\columnwidth]{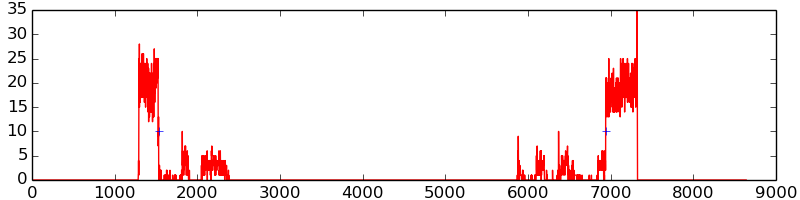}
    \caption{\textbf{Example of distribution and return time detection.} The number of contacts (red line) detected by the sensor before distribution and after return are significantly higher than during the study. By simply using a threshold at $n_c = 10$ contacts per time step, we are able to precisely detect the time this sensor was distributed and returned (blue crosses).}
    \label{fig:data_cleaning}
\end{figure}

Since sensors are continuously powered, we have to identify for each the time window when they are actually in use. In all four conferences, we use the same setup to be able to detect the precise moments when the sensor is handed over to the participant and returned to us (similar setup as in \cite{kontro:EducSci2020}). First, all sensors to be distributed are kept together; thus, the number of contacts detected by sensors before distribution is exceptionally high, immediately recognisable from a normal contact detection situation. Second, extra sensors were added as a security to be used as beacons: as long as a sensor detects them, it indicates that it was not distributed. Finally, we use a similar trick for when the sensors are returned: beacons were added to the returning box, allowing for the precise detection of the returning moment (see Fig.~\ref{fig:data_cleaning}).

In practical terms, we do not distribute a set of name tags and list their identifiers as beacons. We leave a sufficiently large number of beacons in the returning box, allowing us to detect distribution and return times based on the jumps in the number of contacts detected by each sensor. For each sensor, we delete all contacts recorded before distribution and after the return. Finally, all sensors that were not used in contact detection---beacons and undistributed name tags---are removed from the data.

\subsubsection{Data formatting}

After preprocessing and cleaning, the resulting data is a \emph{temporal network} in which the nodes are the participants, and the links represent contacts, appearing and disappearing as time passes. The contact data is formatted as $tij$ file (see Fig.~\ref{fig:tij}), in which each line is a contact occurring at time $t$ between nodes $i$ and $j$. Times are given as UNIX Epoch time.

\begin{figure}[h]
\centering
\begin{minipage}{.35\columnwidth}
\begin{verbatim}
1480486100	89	79
1480486100	35	79
1480486100	56	91
1480486120	56	79
1480486160	89	79
1480486180	56	18
1480486200	56	18
1480486200	35	79
1480486220	56	18
1480486240	56	18
\end{verbatim}
\end{minipage}
\caption{\textbf{Example of a $tij$ file.} This example from the WS16 dataset lists the first 10 contacts recorded, occurring between time 1480486100 and 1480486240. The first line indicates that the contact occurred between participants 89 and 79 at time 1480486100.}
\label{fig:tij}
\end{figure}   

\subsection{Surveys}

\subsubsection{Organisation \& Data anonymity}

In addition to the contact data, we used surveys to gather information about the participants. These self-administered online surveys were available at the beginning of the first day of the conferences. Participants were asked to complete them as soon as possible and typically completed them upon arrival at the venue or within a few hours after their arrival.

\begin{figure}
    \centering
    \includegraphics[width=\columnwidth]{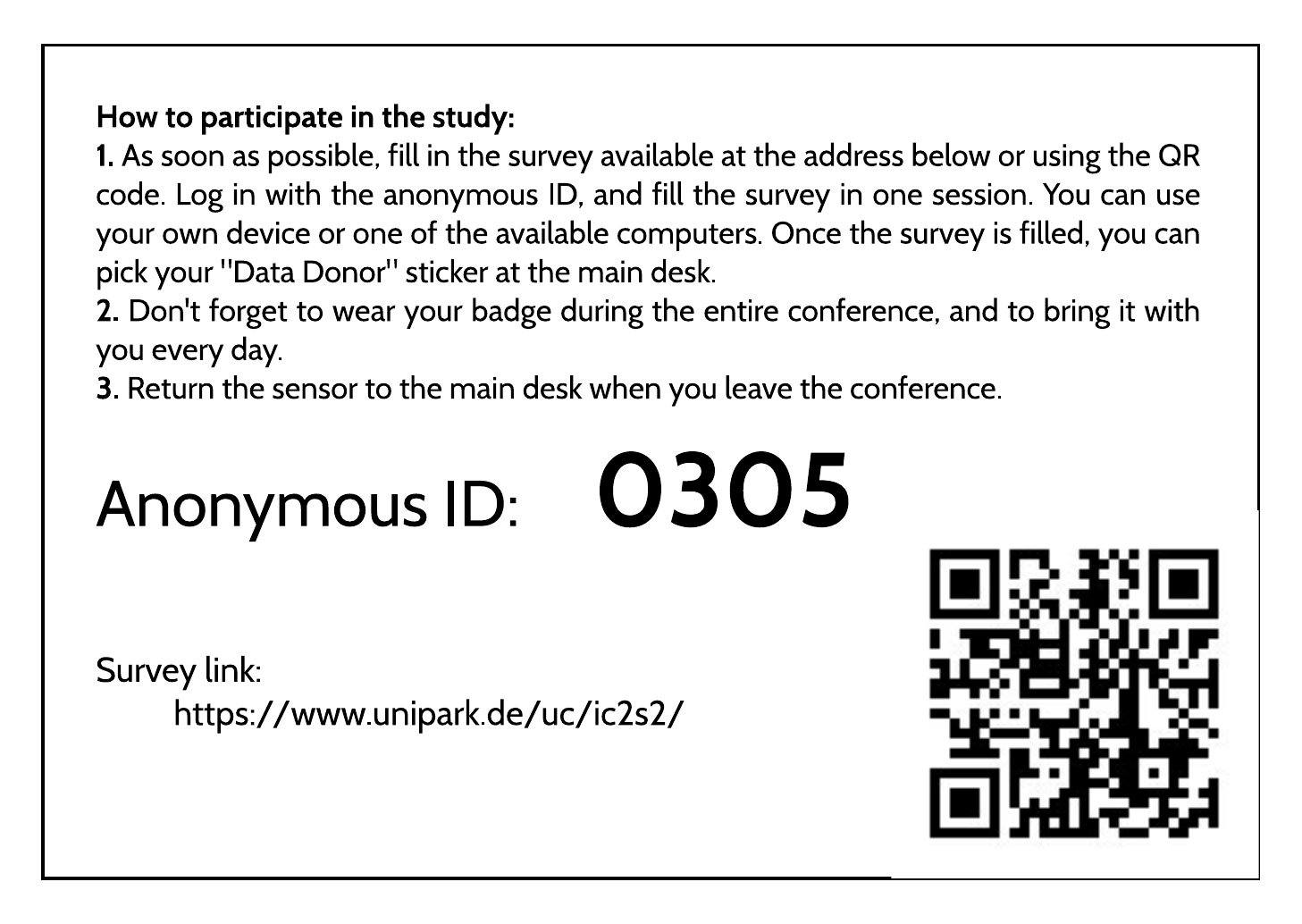}
    \caption{\textbf{Example of survey filling information.} When handed over to participants, each name tag contains an envelope with the information to fill out the survey, the anonymous ID used to connect the survey data to the contact dat, and the link to the survey both written and as a QR code.}
    \label{fig:ID}
\end{figure}

To link the contact data with the survey data while ensuring anonymity, we used a system of anonymous identifiers (IDs). Each sensor has its ID consisting of four numerical digits, which uniquely identify it in the contact data. Along with the name tag, each participant was given an envelope containing this identifier to be used as their identifier when answering the survey (see Fig.~\ref{fig:ID}). Because this anonymous identifier (in the envelopes and sensors) does not have any personal information, we ensure the anonymity of the participants. The anonymous IDs were further replaced by random numbers in the final data, ensuring that no link between the data collection and the final data could be established.

\subsubsection{Content}

The surveys consisted of several sections, covering different axes of inquiry that are relevant to personality science: respondents' sociodemographic characteristics (broadly defined and also including, for example, their disciplinary background and roles at the conference), personality traits (Big Five model~\citep{john2008}), situational perceptions (DIAMONDS model~\citep{rauthmann2014}), scientific attractiveness, motivations to attend the event and perception gap regarding the gender distribution of the crowd. Table~\ref{tab:axes} summarises the content of the survey for each conference.

\begin{table}[t]
    \centering
    \begin{tabular}{|l|c|c|c|c|}
    \hline
    \bf Axis & \bf \rotatebox{90}{WS16} & \bf \rotatebox{90}{ICCSS17} & \bf \rotatebox{90}{ECSS18} & \bf \rotatebox{90}{ECIR19} \\
    \hline
    \bf Sociodemographic characteristics  & x & x & x & x \\
    Age group               & x & x & x & x \\
    Gender                  & x & x & x & x \\
    Age of the oldest child &   &   &   & x \\
    Country of residence    & x & x &   &   \\
    Primary language        & x & x & x & x \\
    Academic status         & x & x & x & x \\
    Disciplinary background & x & x & x & x \\
    Role in the conference  & x & x & x & x \\
    Participation to a previous conference & x & x & x & x \\
    Participation to the pre-symposium     &   &   & x & x \\
    Lunch choice            &   &   &   & x \\
    Number of known persons at the conference &   & x & x & x \\
    \hline
    \bf Big Five personality traits & x & x & x & x \\
    Personality facets & x & x &   &   \\
    \hline
    \bf Situational perceptions (DIAMONDS) &   &   & x & x \\
    \hline
    \bf Scientific attractiveness &   & x & x & x \\
    Self rated attractiveness &   & x & x & x \\
    Number of citations (personal)            &   & x & x & x \\
    Number of citations (other participants)  &   &   &   & x \\
    Number of citations (closest peers)       &   &   &   & x \\
    \hline
    \bf Motivations &   &   & x & x \\
    \hline
    \bf Perception gap                    &   &   & x & x \\
    Share of female participants          &   &   & x & x \\
    Share of professors                   &   &   & x & x \\
    Share of participants younger than 30 &   &   &   & x \\
    Share of German-speaking participants &   &   &   & x \\
    \hline
    \end{tabular}
    \caption{\textbf{Axes of study for surveys.}}
    \label{tab:axes}
\end{table}

In all four conferences, we investigated participants' sociodemographic characteristics; however, the list of items was not always the same. We dropped the question about the country of residence after finding it not relevant. After WS16, we added questions about the number of persons in the conference that participants knew before the event and the number of citations, in parallel with scientific attractiveness, to investigate potential mechanisms for connecting behaviour. In the case of ECSS18 and ECIR19, these events had a pre-symposium, so we asked about participation in these activities. Finally, for ECIR only, we added questions about lunch options (for organisation purposes) and the number of citations of other participants and peers to have insight into how participants see themselves concerning the crowd and their peers.

The second part of the study concerns personality traits, which we assessed using the established Big Five model \citep{john2008}. In the first two conferences, we administered the 30-Item BFI-2-S \citep{soto2017}, which allows investigating 15 narrow personality facets in addition to the Big Five domains (Openness, Conscientiousness, Extraversion, Agreeableness, and Negative Emotionality). In later conferences, we opted for shorter Big Five instruments, namely the 15-item BFI-2-XS \citep{soto2017} for ECSS18 and the 10-item BFI-10 \citep{rammstedt2007} for ECIR19, to make space for other items in the survey. The ultra-short BFI-2-XS and BFI-10 allow for an exploration of Big Five domains but not facets. 

To broaden the space of individual-differences constructs assessed, at ECSS18 and ECIR19 we added a measure of situational perceptions as conceived in the DIAMONDS model (Duty, Intellect, Adversity, Mating, pOsitivity, Negativity, Deception, Sociality) \citep{rauthmann2014}. Situational perceptions refer to how people perceive and construe situations, including the situation's action imperatives. To measure these situational perceptions, we slightly adapted the S8-III, an ultra-short scale measuring each of the eight dimensions with one item \citep{rauthmann2015}. We reworded the introduction such that it referred to the specific situation of scientific conferences and slightly changed some items to align them with best practices of item wording.

With the scientific attractiveness axis, we aim to understand whether respondents scientific status is relevant to understanding contact behaviour. Depending on the conference, we assessed scientific attractiveness in terms of perceived status but also several factual measures such as number of citations.

The motivations axis contains a simple question about the participant's motivations to attend the conference. This axis complements personality traits and the more generic DIAMONDS situation perceptions: it aims at understanding whether behaviour in such contexts is more directed by the nature of the participants or by their intentions.

Finally, the perception gap axis gathers questions about how the structure of the crowd is perceived by the participants in terms of the size of minorities/majorities. This information can inform us about disparities in perception, which can then be correlated with the social network structure as given by the contact data.

For a detailed description of the questions for each survey, the codebooks and questionnaires of the surveys are available with the contact data (See \ref{sec:Data}).

\subsection{Transparency, Openness, and Reproducibility}\label{sec:Data}

\textbf{Pre-registration:} The studies are exploratory and thus were not pre-registered.

\textbf{Hypothesis Testing:} The aim of the present paper is only to present the collected data and does not test any hypothesis.

\textbf{Data:} The contact data are available in GESIS's SowiDataNet$\vert$datorium at the following link:

\url{https://doi.org/10.7802/2351}.

For privacy reasons, the raw contact data are not available, as it contains the sensor IDs that were used during the data collection. For privacy reasons and to comply with the legal regulations concerning the collecting, use and sharing of personal data (GDPR), the complete survey data are available only through direct request to Mathieu G\'enois (mathieu.genois@cpt.univ-mrs.fr). The sharing of these data requires the signature of a sharing agreement that imposes several restrictions, in order to prevent inappropriate uses of the data. An excerpt of the survey data are however available along with codebooks, questionnaires and forms of consent at the following link:

\url{https://doi.org/10.7802/2352}.

This excerpt contains the information about Age class and Gender for WS16 and ICCSS17, Age class only for ECSS18 and ECIR19.

In order to comply with legal regulations about data use, access to the contact data and the survey data excerpt is restricted to scientific purposes only.

\textbf{Scripts, Code, Syntax:} The code for the extracting and preprocessing of the raw data gathered by the antennas is not available, for proprietary reasons. The program to produce Table~\ref{tab:contact_properties} and Figures~\ref{fig:activity_timelines} to \ref{fig:degree} is available at:

\url{https://mycore.core-cloud.net/index.php/s/6J1tgSnubq9imYG}.

It relies on the \verb,tempnet, library available at:

\url{https://github.com/mgenois/RandTempNet}.

\textbf{Other Supplements:} No supplementary information is provided with this paper.

\section{Results}

This section presents general statistics of the data sets.

\subsection{General description of the events}

The data were collected during four events organised by GESIS, the Leibniz Institute for Social Sciences, in Cologne, Germany. Throughout the manuscript, we refer to them using the following labels:

\begin{itemize}
\item \textbf{WS16}: the 3rd GESIS Computational Social Science Winter Symposium, held on November 30 and December 1, 2016~\cite{WinterSymposiumURL}. This event was part of a series on computational social science organised by GESIS. This edition had the specific topic of ``Understanding social systems via computational approaches and new kinds of data".

\item \textbf{ICCSS17}: the International Conference on Computational Social Science, held from July 10 to 13, 2017~\cite{IC2S2URL}. Broadly speaking, the conference is known for bringing interdisciplinary researchers together for advancing social science knowledge through computational methods.

\item \textbf{ECSS18}: the Eurosymposium on Computational Social Science, held from December 5 to 7, 2018~\cite{EuroSymposiumURL}. This event was part of the European Symposium Series on Societal Challenges in Computational Social Science. This edition had the headline of ``Bias and Discrimination''. 

\item \textbf{ECIR19}: the 41st European Conference on Information Retrieval, held from April 14 to 18, 2019~\cite{ECIRURL}. The conference is the European forum for the presentation of research in the field of Information Retrieval.

\end{itemize}

Though all events occurred in Cologne, Germany, they were organised in different locations. WS16 was held at the KOMED convention centre at MediaPark~\cite{KOMEDURL}, whereas ICCSS17, ECSS18, and ECIR19 took place at the Maternushaus hotel~\cite{MaternushausURL}. 

The first three conferences (i.e., WS16, ICCSS17, and ECSS18) were interdisciplinary, gathering researchers from Social Sciences, Computer Sciences and Natural Sciences. In contrast, ECIR19 was focused on Computer Science. For the last three conferences  (i.e., ICCSS17, ECSS18, and ECIR19), the first day consisted of a separate workshop/pre-symposium day, for which contact data was also gathered except for ECSS18, for which we have contact data only for the main conference on December 6 and 7.

\begin{table}[b]
  \centering
  \begin{tabular}{ccccc}
    \toprule
    Study  & WS16           & ICCSS17        & ECSS18         & ECIR19\\\midrule
    $N$    & 149            & 339            & 211            & 270\\
    $N_p $ & 144 (96.6\,\%) & 284 (83.8\,\%) & 205 (97.2\,\%) & 190 (70.3\,\%)\\
    $N_p^* $ & 144 (96.6\,\%) & 277 (81.7\,\%) & 171 (81.0\,\%) & 178 (65.9\,\%)\\\midrule
    $N_c$  & 138 (95.8\,\%) & 274 (96.5\,\%) & 164 (80.0\,\%) & 172 (90.5\,\%)\\
    $N_d$  & 122 (83.3\,\%) & 213 (75.0\,\%) & 155 (75.6\,\%) & 140 (73.7\,\%)\\\bottomrule
  \end{tabular}
  \caption{\textbf{Statistics of participation to the studies.} $N$ is the total number of participants; $N_p$ is the number who agreed to take part in the study; $N_p^*$ is the number for which we have data (contact and/or survey); $N_c$ is the number for which we have contact data; $N_d$ is the number for which we have at least partial sociodemographic information. Percentages for $N_p$ and $N_p^*$ are calculated with respect to $N$; percentages for $N_c$ and $N_d$ are calculated for the studied population and thus with respect to $N_p$.}
  \label{tab:stat}
\end{table}

Table~\ref{tab:stat} lists basic statistics of participation to the studies. Overall, we have a very high participation rate, with more than 70\,\% of attendees partaking in the studies. In the case of contact data, we have excellent coverage of the conferences' crowds; it is greater than 90\,\% for three studies and 80\,\% for ECSS18. The survey response rate is also good. We have at least partial information for more than 70\,\% of the studied population.

\subsection{Properties of the contact networks}

\begin{table}[t]
    \centering
  \begin{tabular}{ccccc}
    \toprule
           & WS16     & ICCSS17  & ECSS18  & ECIR19   \\\midrule
    $C$    & 153\,371 & 229\,536 & 96\,362 & 132\,949 \\
    $\rho$ & 0.793    & 0.495    & 0.567   & 0.550 \\
    $\left<k\right>$ & 108.6 & 135.2 &  92.4 &  94.1 \\
    $\left<c\right>$ & 0.868 & 0.694 & 0.717 & 0.746 \\
    \bottomrule 
    \end{tabular}
    \caption{\textbf{General properties of the contact networks.} $C$: total number of instantaneous contacts recorded; $\rho$: density of the aggregated network, \emph{i.e.} the fraction of possible connections that occurred during the event; $\left<k\right>$: average degree of the aggregated network, \emph{i.e.} the average number of persons one participant met during the event; $\left<c\right>$: average clustering of the aggregated network.}
    \label{tab:contact_properties}
\end{table}

\begin{figure*}[t]
    \centering
    \includegraphics[width=\columnwidth]{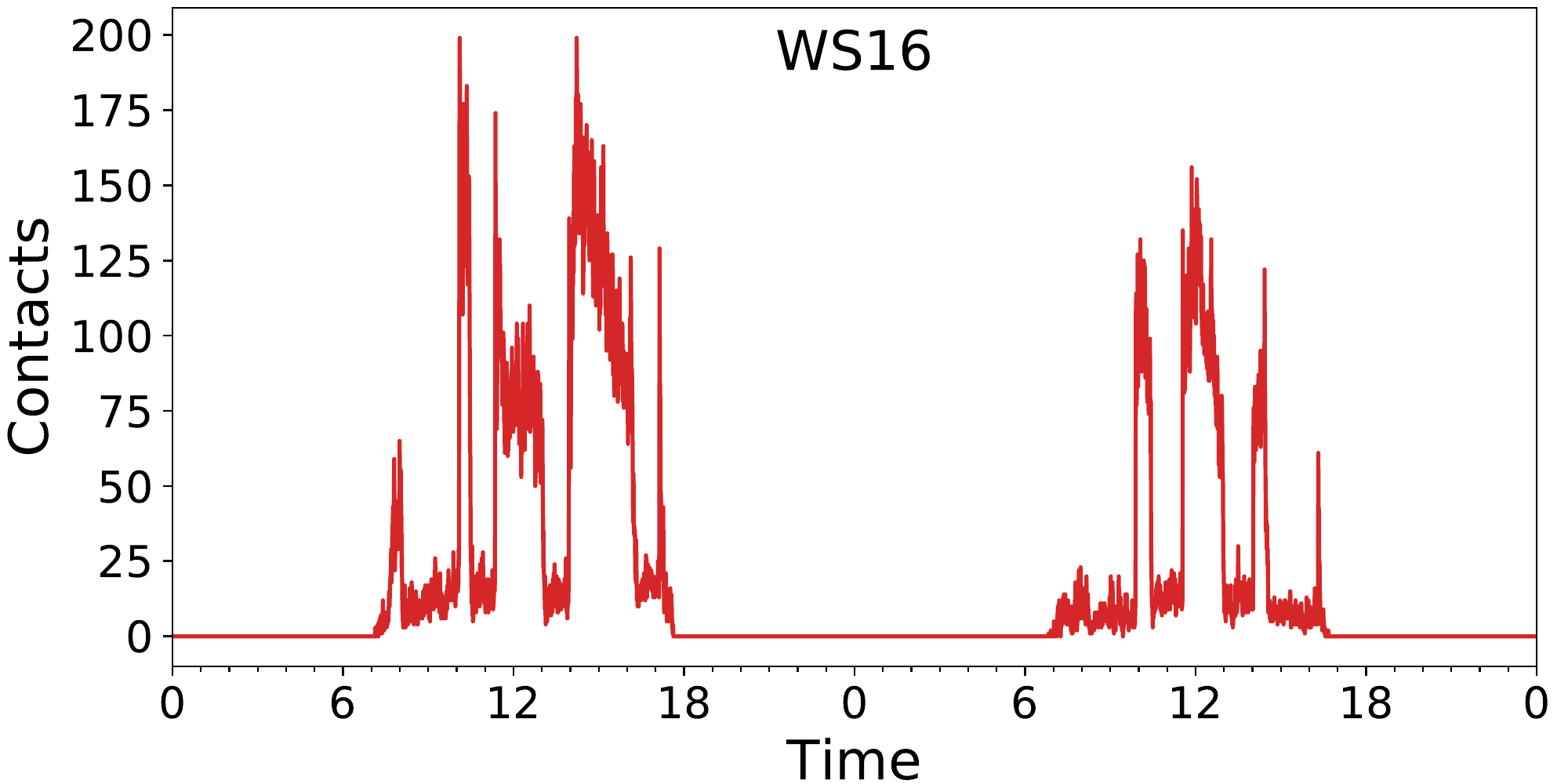}
    \includegraphics[width=\columnwidth]{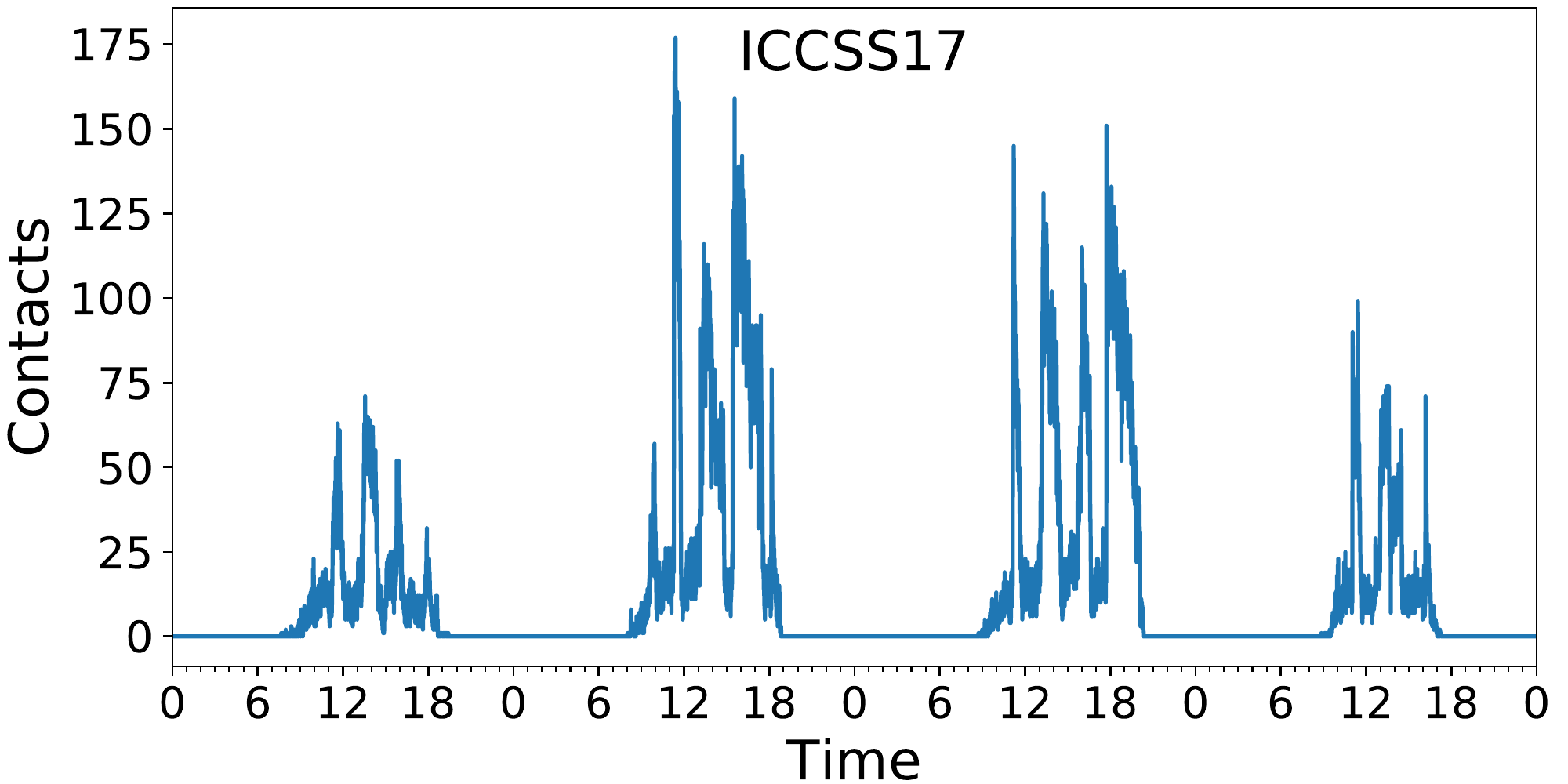}
    \\
    \includegraphics[width=\columnwidth]{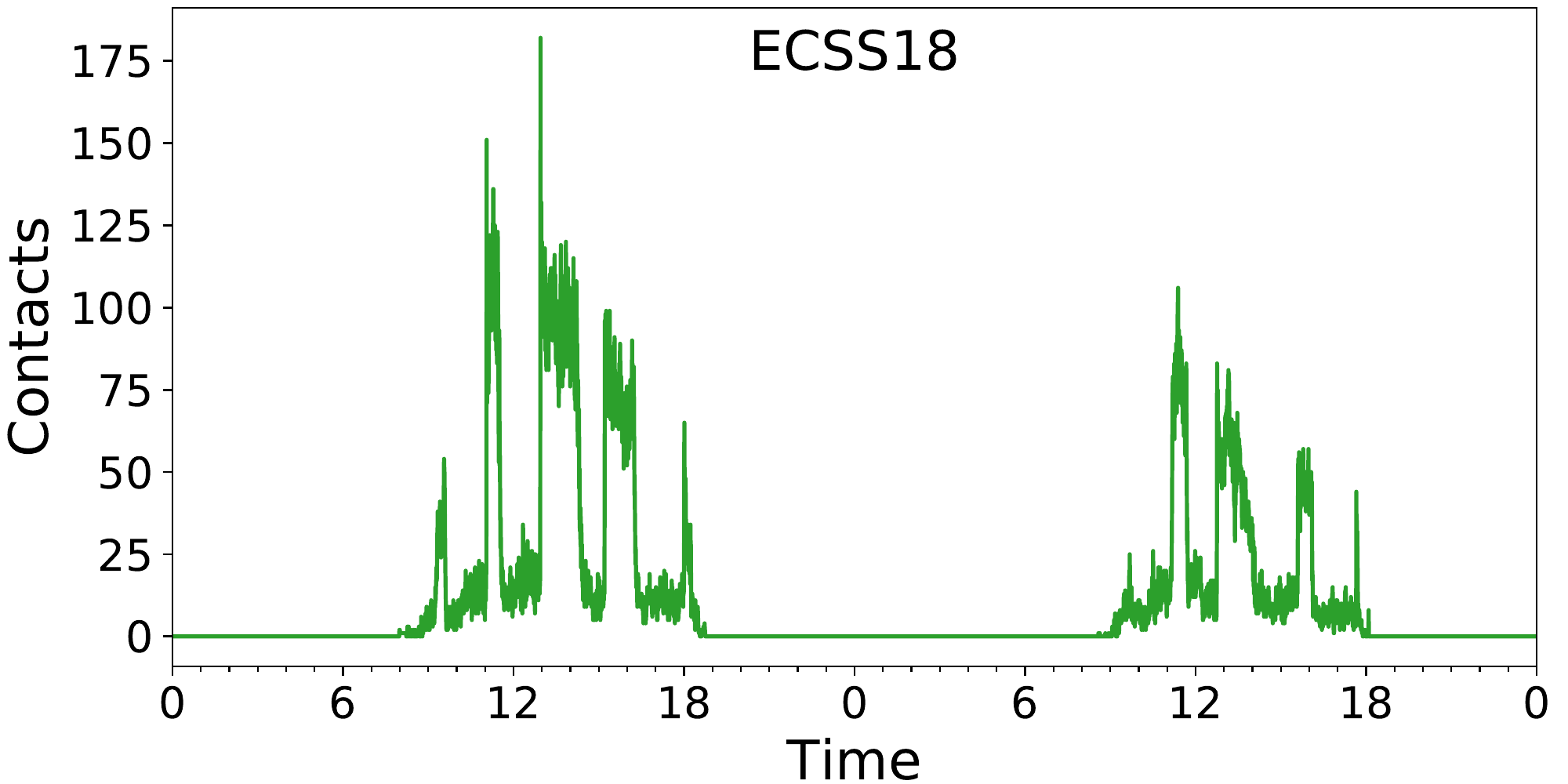}
    \includegraphics[width=\columnwidth]{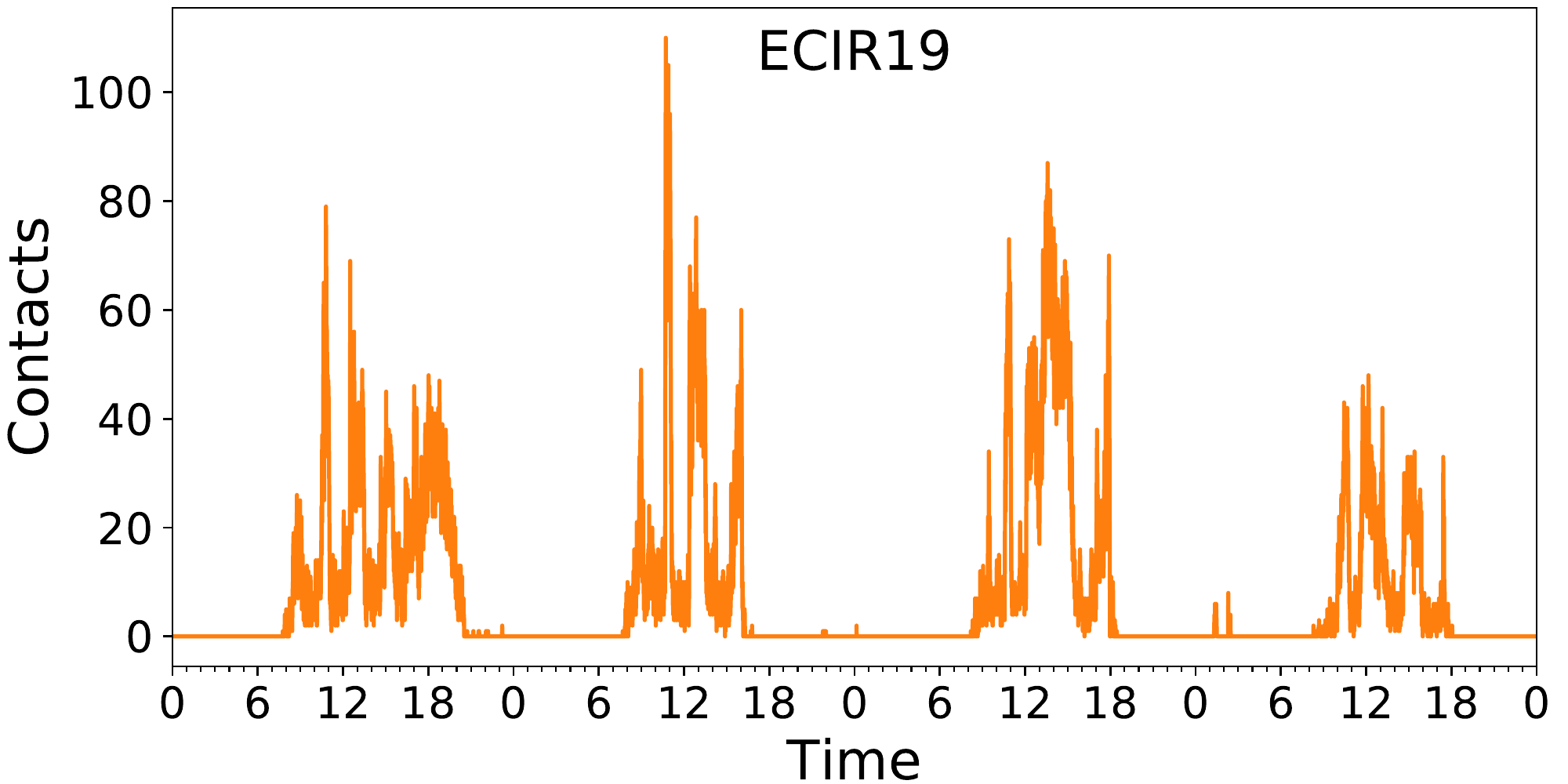}
    \caption{\textbf{Activity timelines.} We plot the total number of contacts occurring in each 20 seconds time step. Curves exhibit the circadian rhythm (activity during the day and no activity at night) and alternating periods of social times (coffee breaks, lunch, poster session) and low activity windows (talk sessions).}
    \label{fig:activity_timelines}
\end{figure*}

\begin{figure}[t]
    \centering
    \includegraphics[width=\columnwidth]{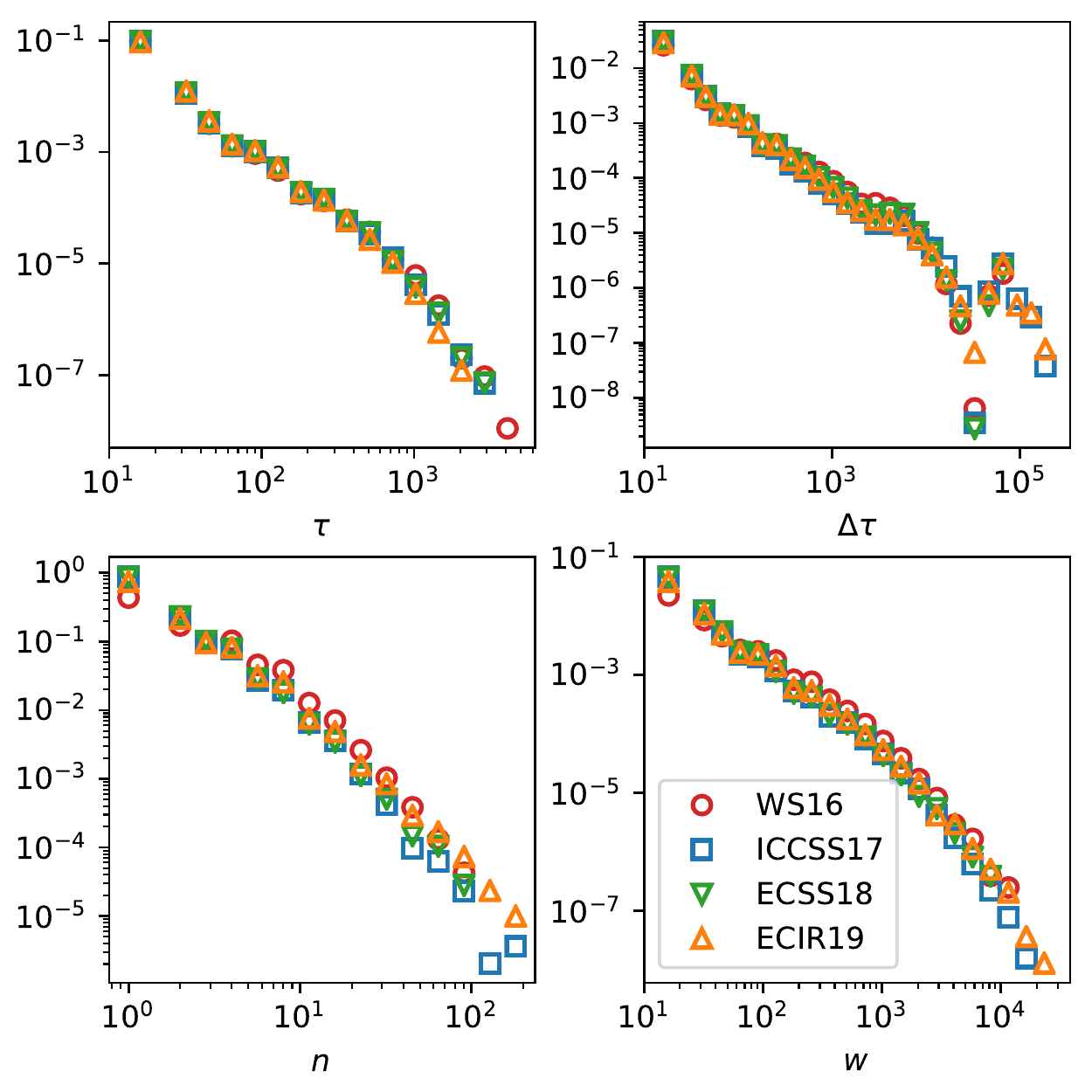}
    \caption{\textbf{Distributions of temporal features.} $\tau$ are the contiguous contact durations; $\Delta\tau$ are the inter-contact durations; $n$ are the number of contacts between each pair of participants; $w$ are the total contact durations for each pair of participants. All distribution have a heavy tail, a typical feature of human behaviour: most of the interactions are short and not repeated, most intervals between interactions are short; at the same time long interactions, long intervals, frequent repetitions and strong links, though rarer, are not negligible. The depletion/inflation feature of the $\Delta\tau$ distribution is due to the circadian rhythm.}
    \label{fig:temporal}
\end{figure}

\begin{figure}[t]
    \centering
    \includegraphics[width=.49\columnwidth]{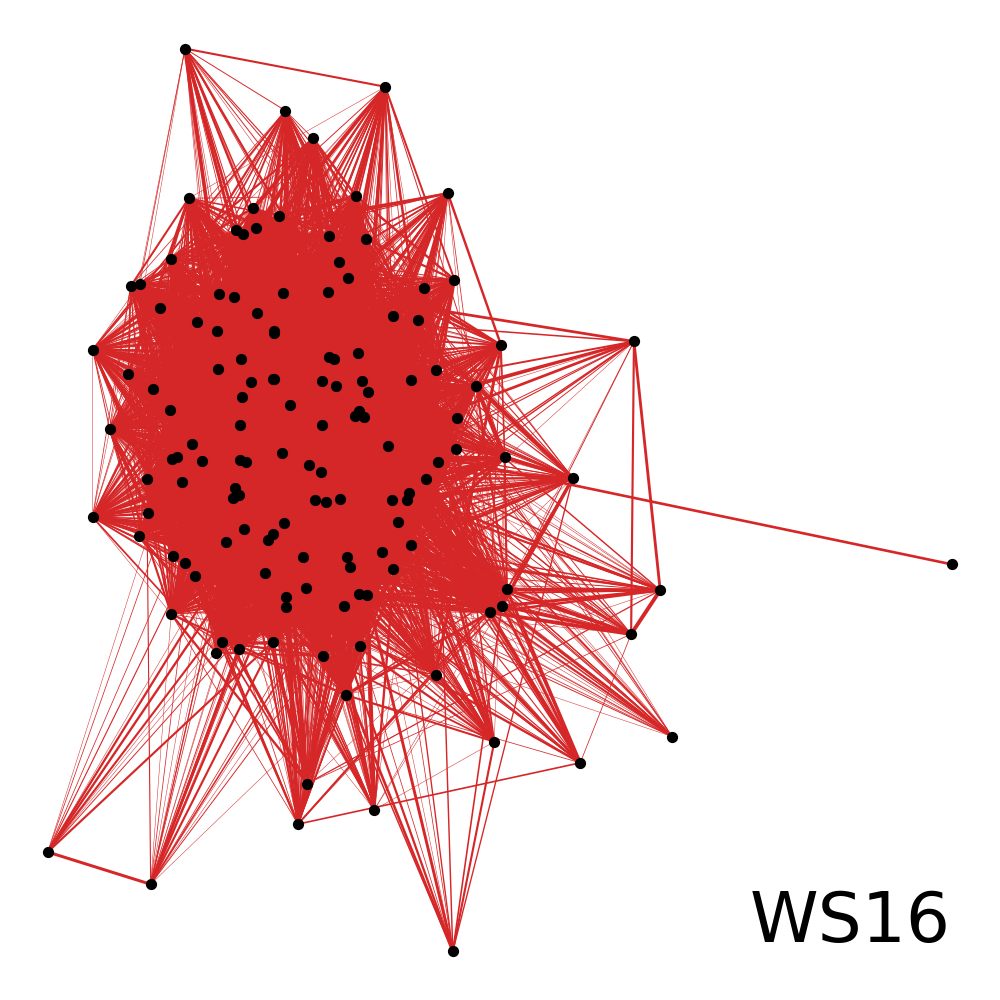}
    \includegraphics[width=.49\columnwidth]{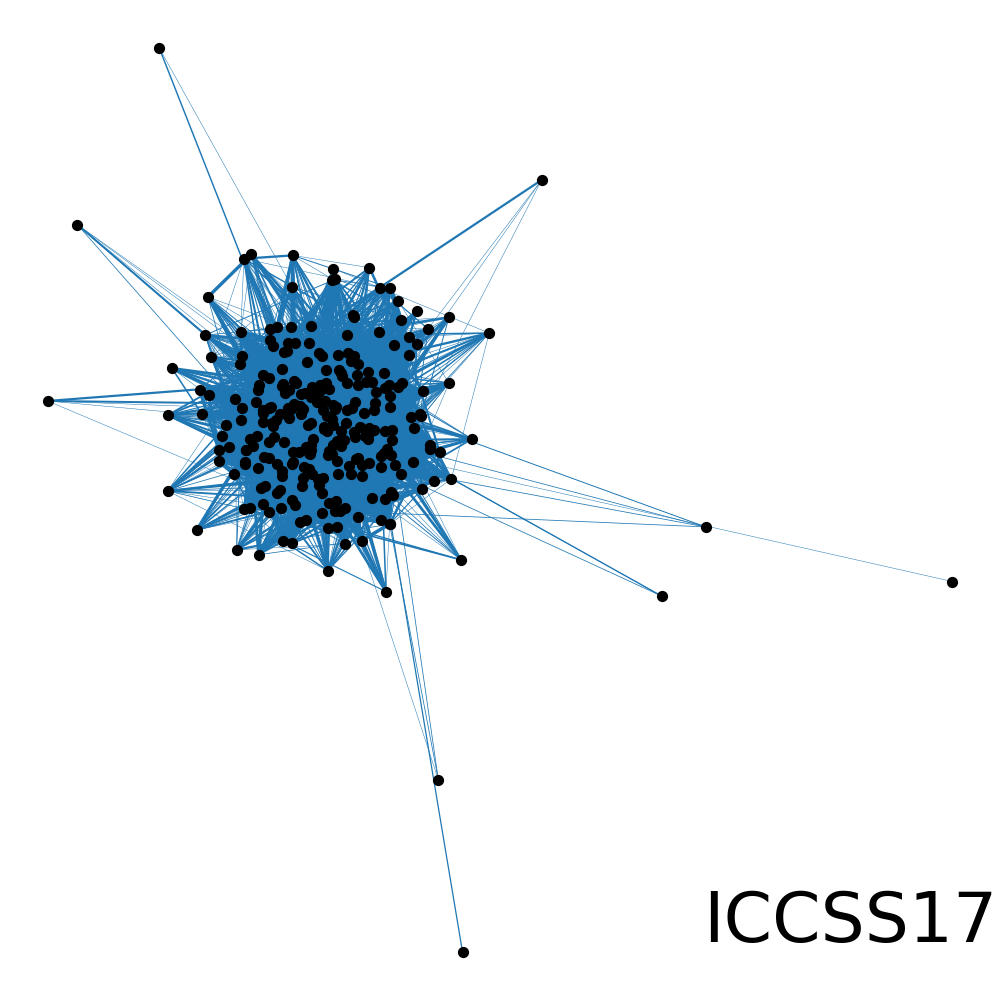}
    \\
    \includegraphics[width=.49\columnwidth]{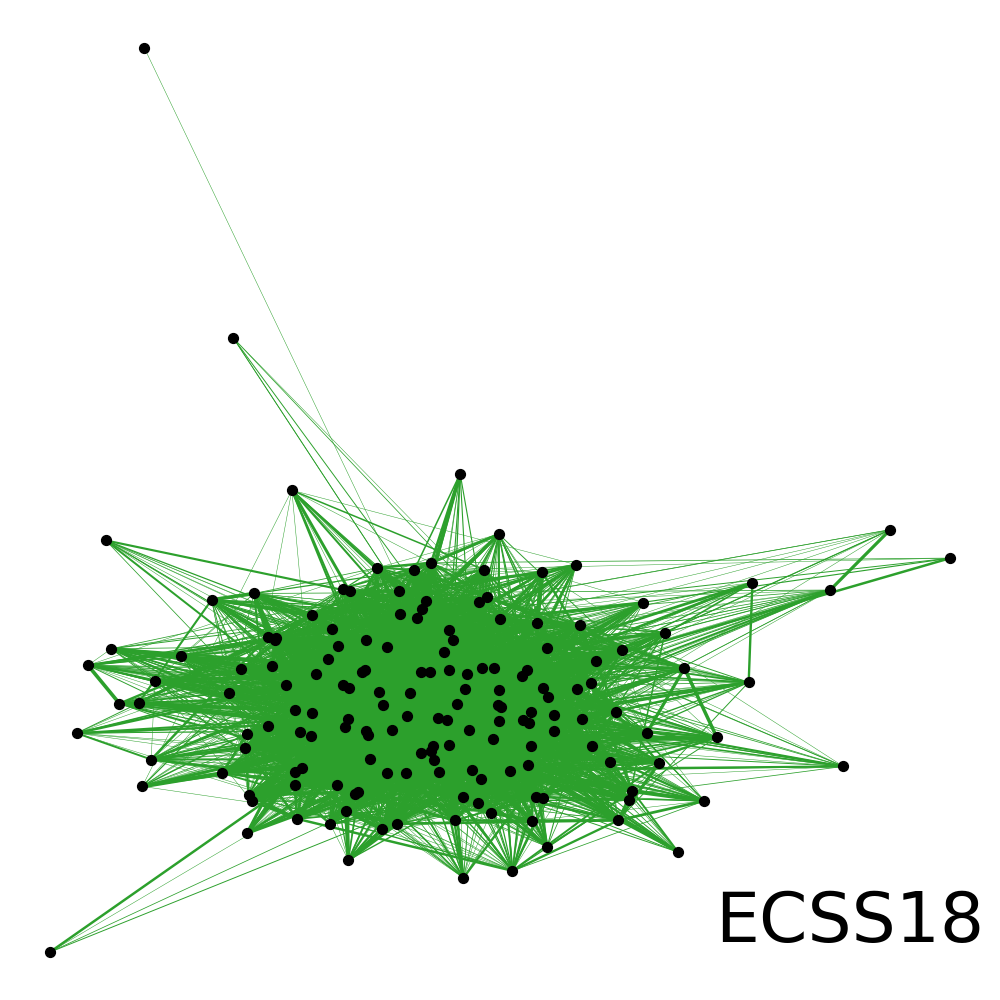}
    \includegraphics[width=.49\columnwidth]{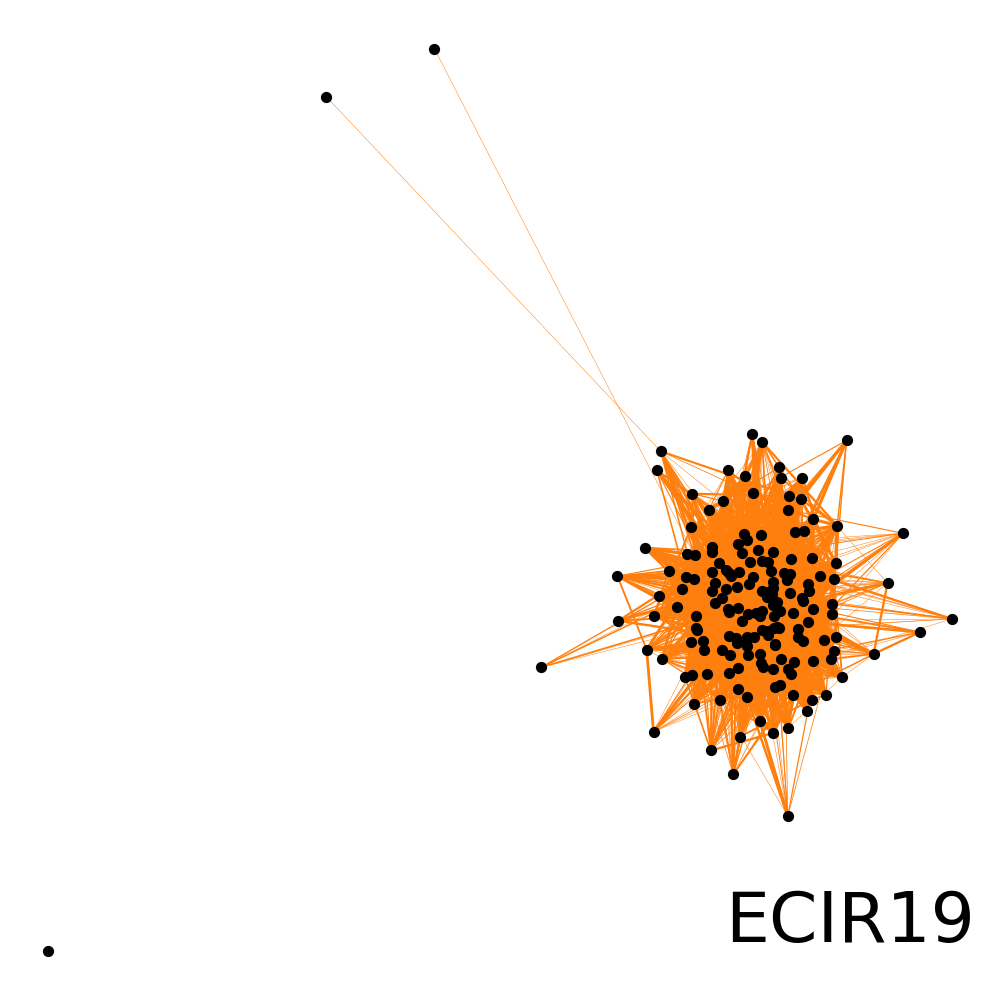}
    \caption{\textbf{Visualisation of the aggregated contact networks.} Nodes are individuals; a link exists between two individuals if they have been at least once in contact during the event; the width of the link is proportional to the total contact duration between the two individuals. Node positions were set using a spring layout, were links are equated to springs with a stiffness proportional to its weight; all networks exhibit the same large number of links and absence of visible global structure.}
    \label{fig:network}
\end{figure}

\begin{figure}[t]
    \centering
    \includegraphics[width=.49\columnwidth]{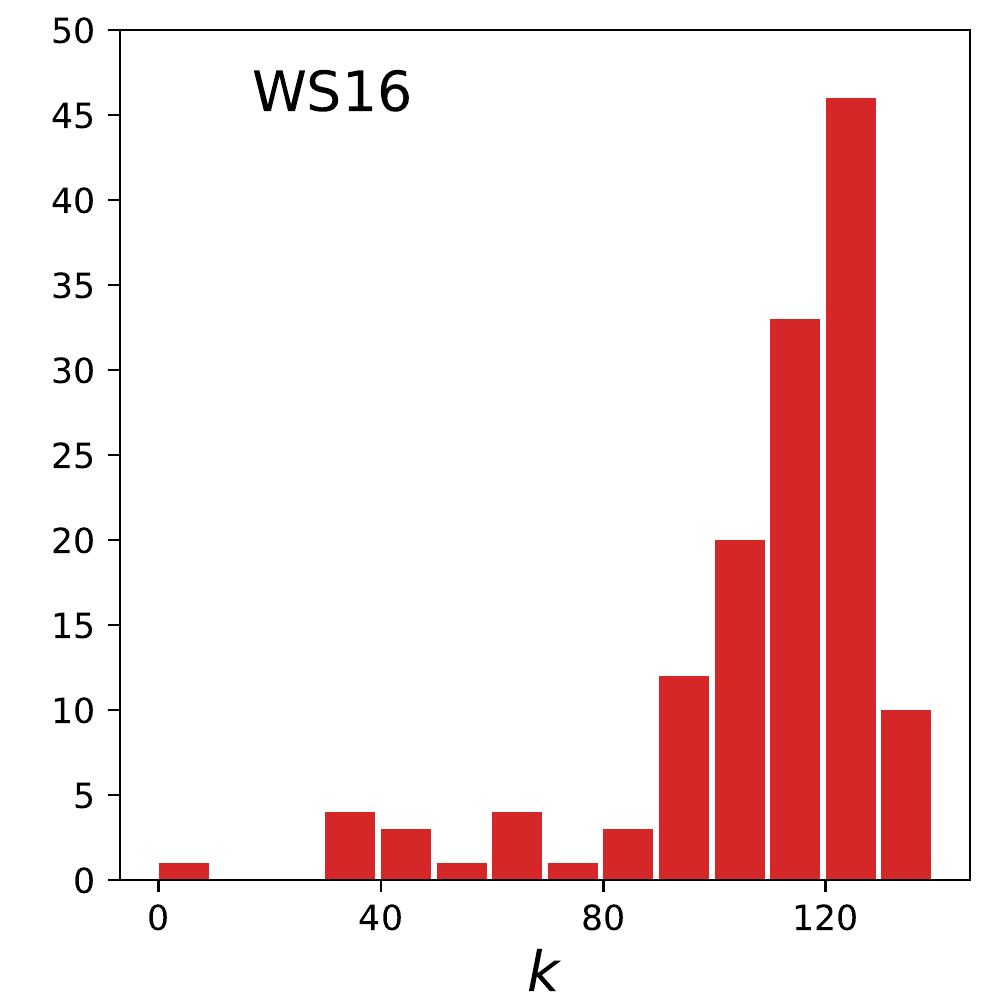}
    \includegraphics[width=.49\columnwidth]{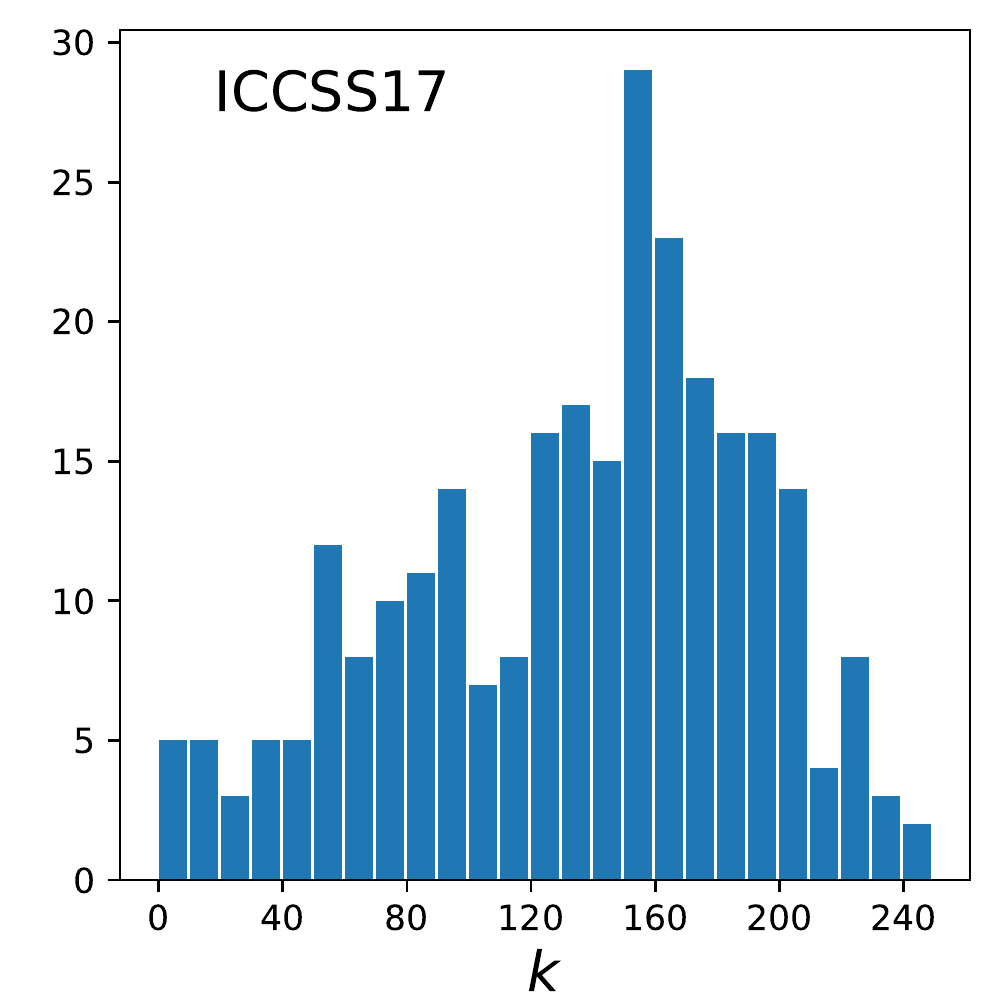}
    \\
    \includegraphics[width=.49\columnwidth]{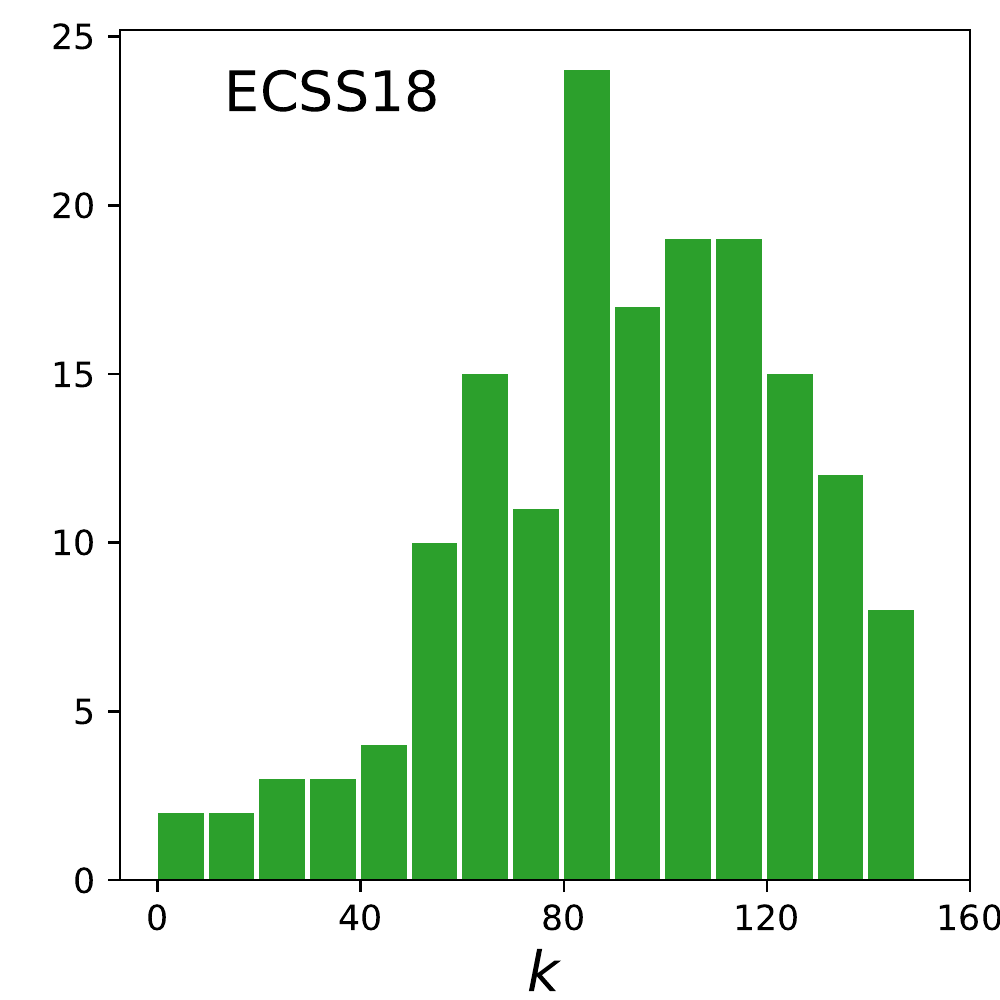}
    \includegraphics[width=.49\columnwidth]{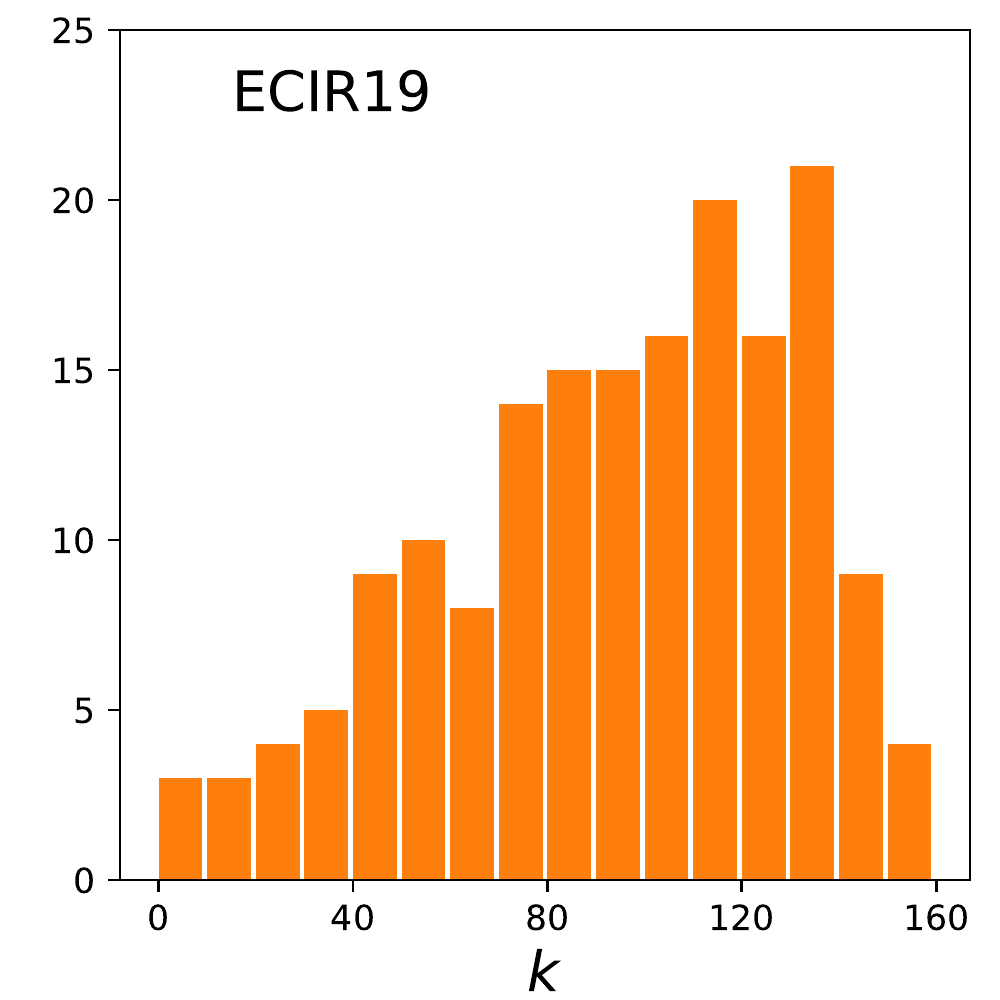}
    \caption{\textbf{Degree distributions.} The degree of an individual is the number of connection it has in the aggregated network, \emph{i.e.} the number of individuals with whom it has interacted at least once during the event. All distributions are skewed towards high values, indicating a large mixing: most people have interacted at least once with a significant fraction of the crowd.}
    \label{fig:degree}
\end{figure}

The temporal networks obtained through the SocioPatterns studies consist of temporal links that indicate, every 20 seconds, which participants are in contact. We denote $C$ as the total number of these instantaneous contacts, which describes the overall recorded activity in an event (see Table~\ref{tab:contact_properties}). This activity changes over time, so we further define \emph{contact activity} as the number of instantaneous contacts occurring per time step. It describes the evolution of the interaction level between participants (Fig.~\ref{fig:activity_timelines}). This evolution is similar for all conferences: we observe a circadian rhythm, with active days and inactive nights. The active periods exhibit a wave shape with a progressive increase at the beginning and a decrease at the end, modulated by the succession of high and low activity periods. High activity periods are ``social times'' such as registration, coffee/lunch breaks, or poster sessions; low activity periods are talk sessions.

To assess the dynamics of face-to-face interactions, we evaluate some basic statistics regarding the contacts (see Fig.~\ref{fig:temporal}). We define any instantaneous contacts occurring sequentially without in-between gaps as a continuous contact with a duration of $\tau$ (i.e., an interaction). With this definition, we can then explore the overall temporal properties of the interactions (i.e., the distributions of $\tau$). Additionally, we can examine the inter-contact durations, denoted $\Delta\tau$, between two consecutive interactions between the same participants. Furthermore, we evaluate the number of contacts $n$ and the total contact duration (i.e., weight) $w$ occurring between two participants. By examining the empirical distributions of these quantities, we find well-known, large-tail shaped distributions. This finding indicates that the most numerous contacts last 20 seconds, the most numerous inter-contact durations last 20 seconds, most pairs of participants interacted only once, and for one contact of 20 seconds only. However, extremely long instances of each of these properties also occur, with a small but not negligible probability, as indicated by the roughly power-law aspect of the distributions. Finally, the distribution of $\Delta\tau$ exhibits the usual depletion/inflation feature caused by the circadian rhythm in the activity data. 

By flattening the temporal network across the temporal dimension, we obtain an \emph{aggregated network} in which nodes are the participants, and a link exists between two nodes if the participants have interacted at least once during the event. We perform a standard analysis of these networks (see Table~\ref{tab:contact_properties}).

We first computed the \emph{density} $\rho$ of the aggregated network, \emph{i.e.,} the fraction of links that exist in the network with respect to all possible links. We find that in all four studies the aggregated networks are very dense. This finding is primarily because the venues were somewhat crowded, ensuring that each participant came into contact with a significant fraction of the rest of the crowd. One can indeed see on visualisations of the networks that connections are very numerous (Fig.~\ref{fig:network}).

The degree $k$ of a node is the number of links it has in the network, indicating the number of participants the individual has been at least once in contact. The high density of the networks appears on the average degree as well as the degree distributions, which are skewed towards high values. This shows that, indeed, most participants interacted at least once with most of the other participants (Fig.~\ref{fig:degree}).

The clustering $c$ of a node is a measure of the connectedness of its neighbours. A high value of $c$ indicates that most of the persons with whom one participant has interacted also interacted with each other; a low value of $c$ indicates the opposite. We computed the \emph{average clustering} of the networks, and find the values to be very high, which is also a consequence of the high densities.

\subsection{Survey information}

The accompanying surveys assessing the axes shown in Table~\ref{tab:axes} were conducted as online surveys but administered on-site. After arriving at the conferences, participants were invited to participate in the survey, which they could fill out on laptops provided by the conference organisers or on their own devices. For linkage of the survey data to the sensor data, the first item of each survey always required participants to provide their sensor ID. 

At WS16 and ICSS17, there was only one survey. Because some of the survey items might be reactive (i.e., respond to the experiences at the conference), efforts were made to encourage participants to fill in the survey immediately after arriving at the conference---the majority of participants filled in the survey on the first conference day. At ECSS18 and ECIR19, participants were invited to participate in a second survey toward the end of the conference, in which additional questions that depended on participants' experiences during the conferences (especially about perception gap) were asked. 

The survey participation rates relative to the number of participants who wore a sensor ranged from 73.7\,\% in ECIR19 to 83.3\,\% in WS16, as shown in Table~\ref{tab:stat}. Item missingness among those who started the survey was negligible (typically $<$5\,\%) at all conferences. The length of the surveys was kept short to avoid interfering with other conference activities and to minimise respondent burden. Respondents typically took between 5-10 minutes to complete the surveys. Median completing times were 5.45 min. for WS16, 7.12 min. for ICCSS17, 5.18 min for ECSS18, and 8.17 min. for ECIR19. The second surveys conducted at ECSS18 and ECIR19 were shorter, with median completion times of 0.87 and 1.30 min, respectively.

\section{Discussion}

The data presented here covers many aspects of social behaviour and individual difference constructs relevant to personality science. Its main advantage is the parallel collection of quantitative data about social interactions and survey data about the individuals, which allows for an exploration of the linkage between a person's characteristics and their social behaviour as measured by the sensors. Furthermore, we present not only one but four data sets collected using the same protocol, making it possible to check for the replicability and reproducibility of phenomena across events.

Among the many possible research questions that can be addressed with this data, we are currently working on two. First, we are exploring the relationship between sociodemographic characteristics and social interactions. We wonder whether different sociodemographic groups exhibit consistent variation in the number of connections they establish and their intensity. Second, we investigate the predictive power of personality traits as defined by the Big Five model for the social behaviour participants exhibit at the conferences. Our ongoing studies are mainly meant to showcase the type of research questions that can be addressed with these data, yet they use only a tiny share of these data's potential. Therefore, we invite other personality scientists to use these data for secondary analyses to explore individual differences in behaviour and their origins.  

\bibliography{biblio}

\clearpage
\onecolumngrid
\appendix

\section{Venue plans}\label{app:plans}

\begin{figure*}[h]
    \centering
    \includegraphics[width=.48\columnwidth]{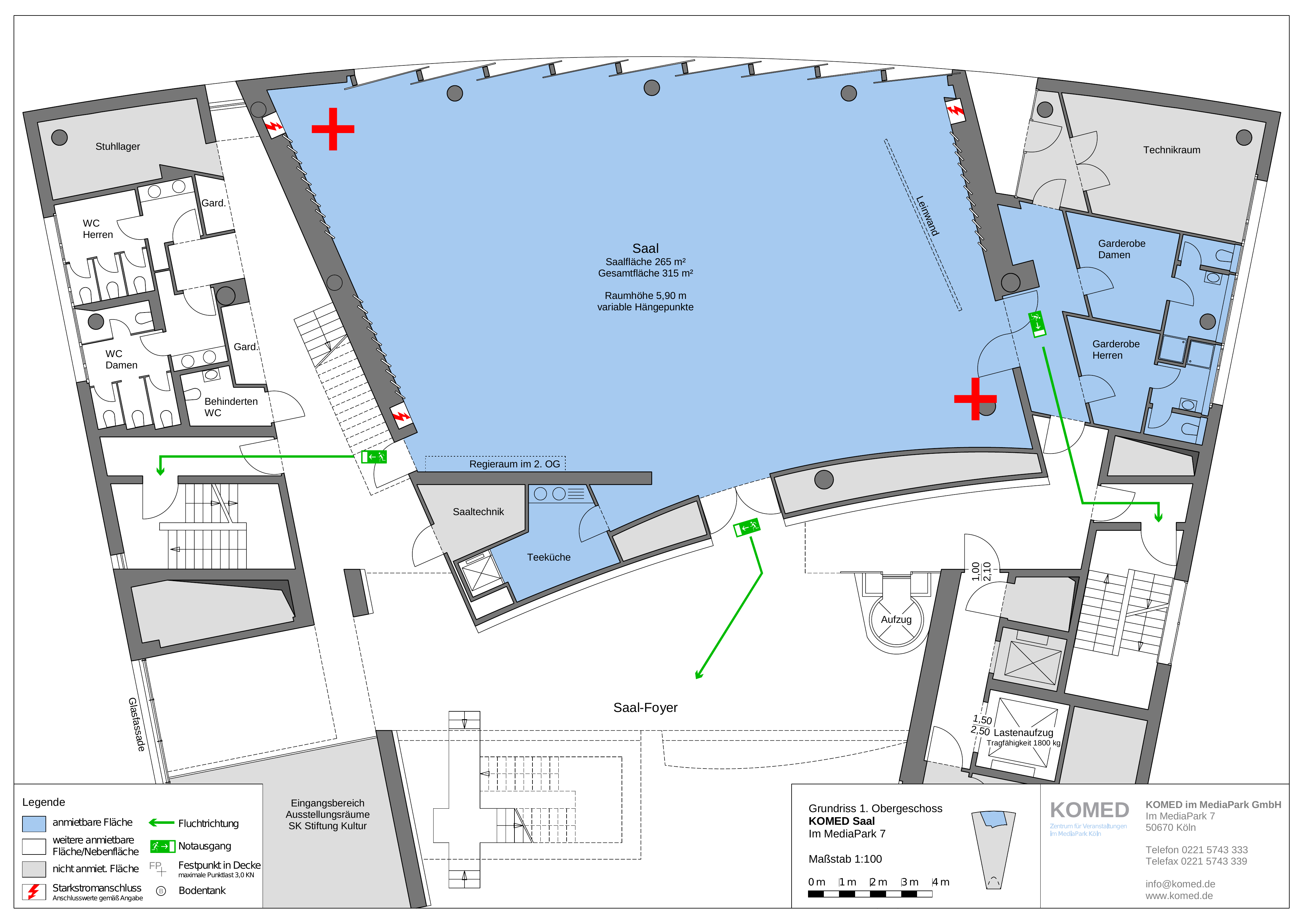}
    \includegraphics[width=.48\columnwidth]{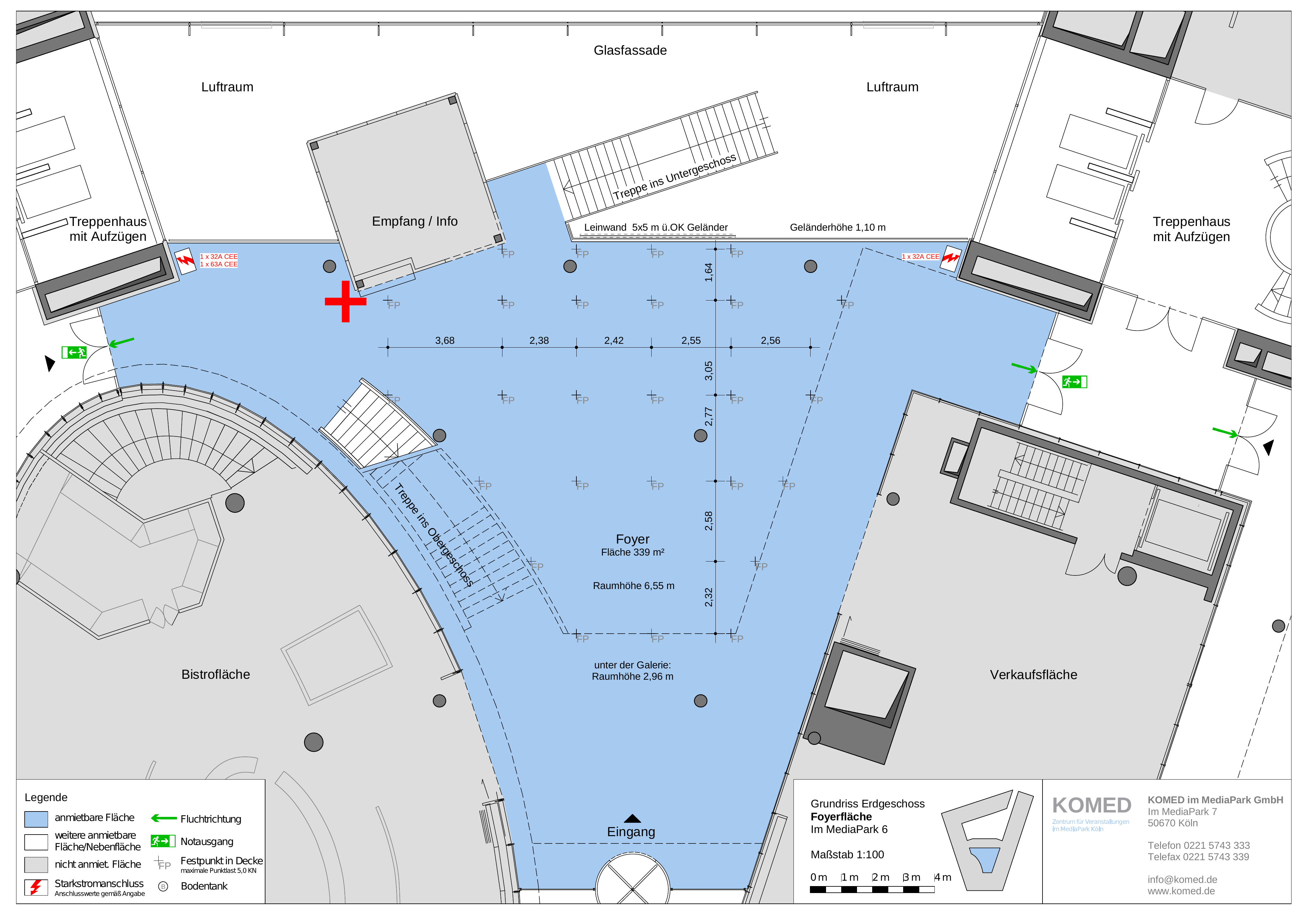}
    \\
    \includegraphics[width=.48\columnwidth]{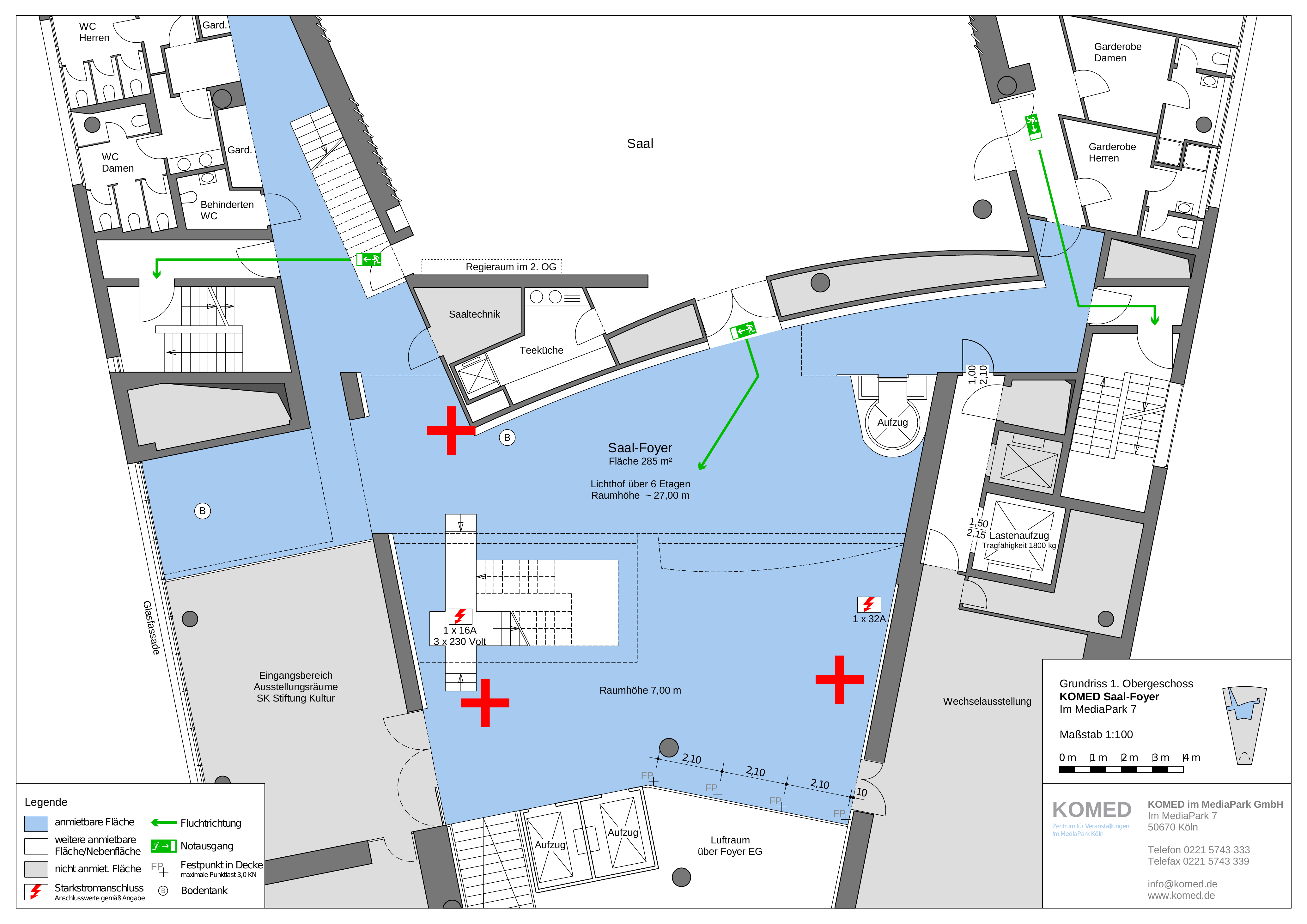}
    \includegraphics[width=.48\columnwidth]{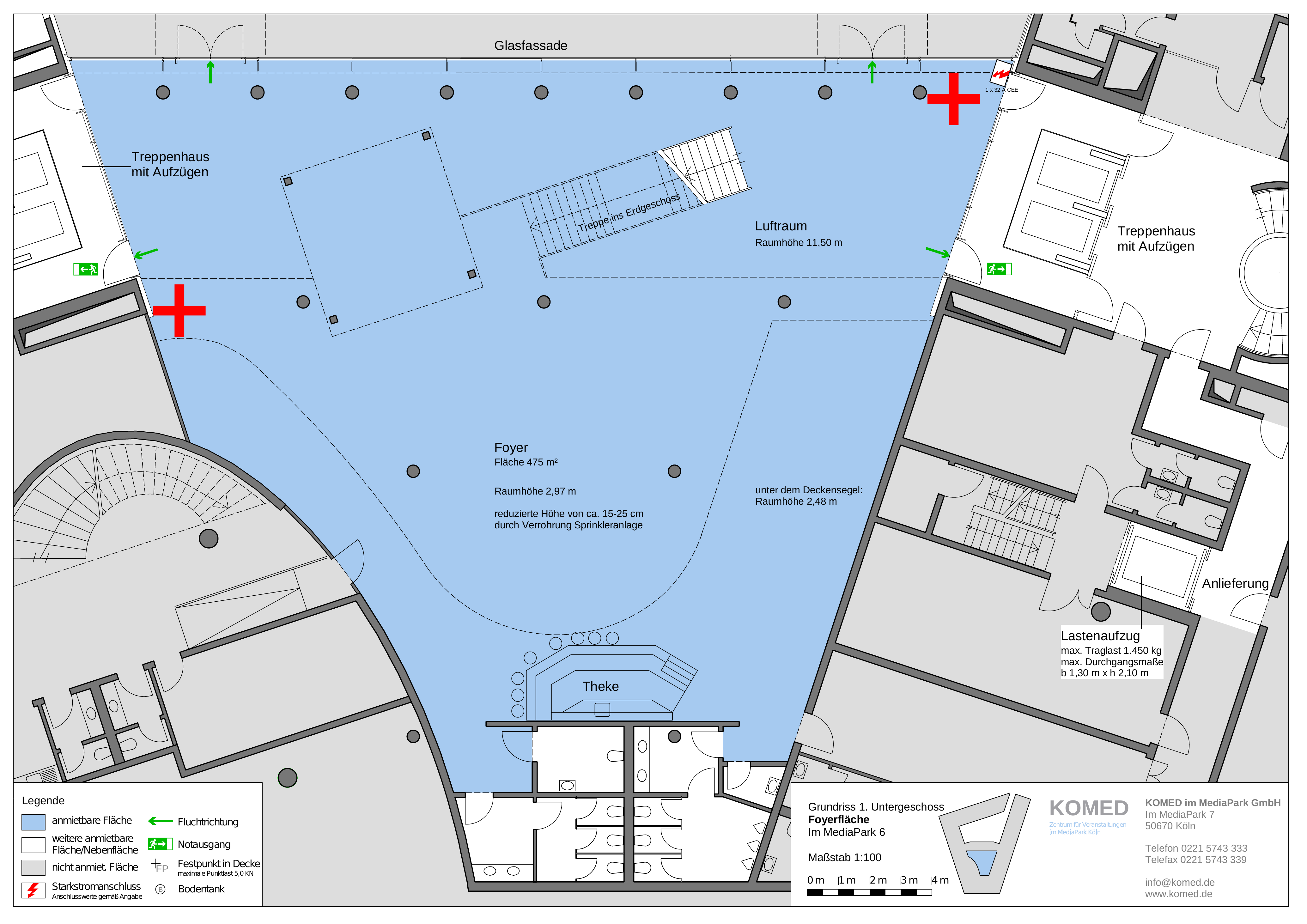}
    \caption{\textbf{Antenna locations for the WS16 conference.} WS16 was organised in the KOMED building in MediaPark in Cologne, Germany. The venue consisted of two separate locations: the conference venue (left column) with the main room (top) and the reception (bottom), and the social venue (right column) on two floors (top: ground floor, bottom: basement). Antennas are indicated by the red crosses. Basic floor plan by KOMED im MediaPark GmbH.}
    \label{fig:plans_WS16}
\end{figure*}

\begin{figure*}[h]
    \centering
    \includegraphics[width=.48\columnwidth]{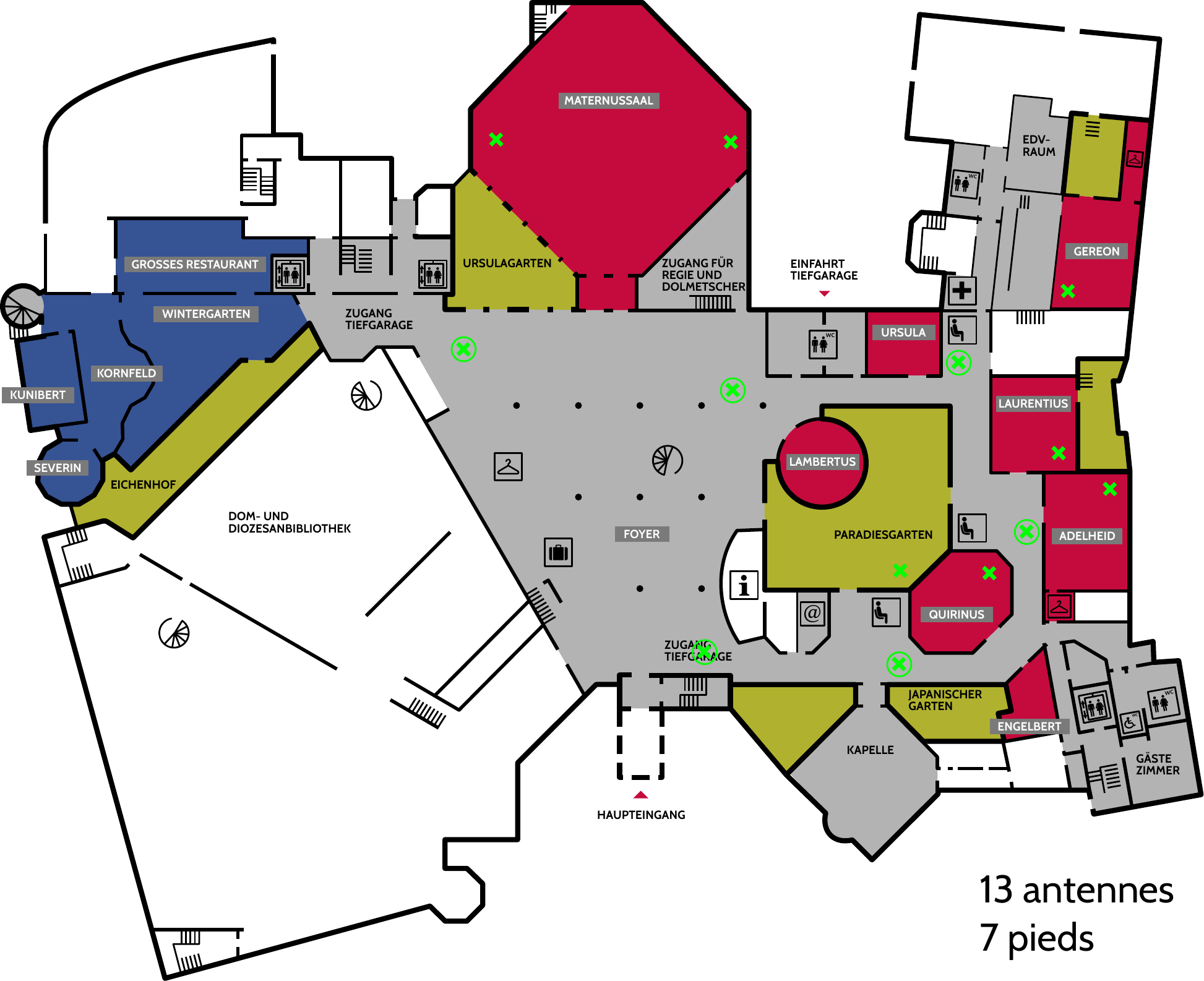}
    \includegraphics[width=.48\columnwidth]{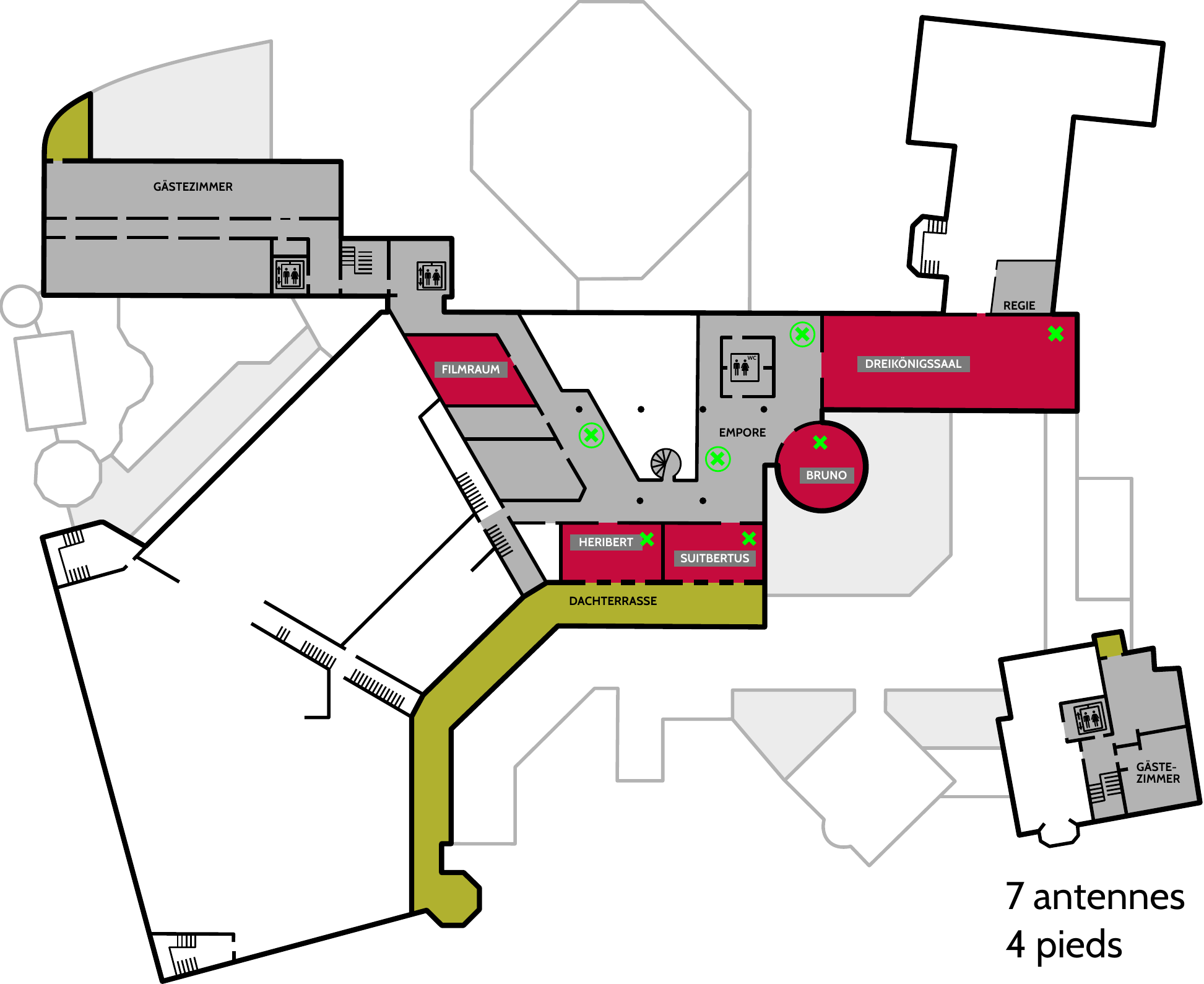}
    \caption{\textbf{Antenna locations for the ICCSS17 conference.} ICCSS17 was organised in the Maternushaus in Cologne, Germany. The venue was organised in two floors (left: ground floor, right: first floor). On the plans, dark red indicate presentation rooms, gray indicate circulation areas, green outside areas. Antennas are indicated by the green crosses. Basic floor plan by Maternushaus.}
    \label{fig:plans_ICCSS17}
\end{figure*}

\begin{figure*}[h]
    \centering
    \includegraphics[width=.48\columnwidth]{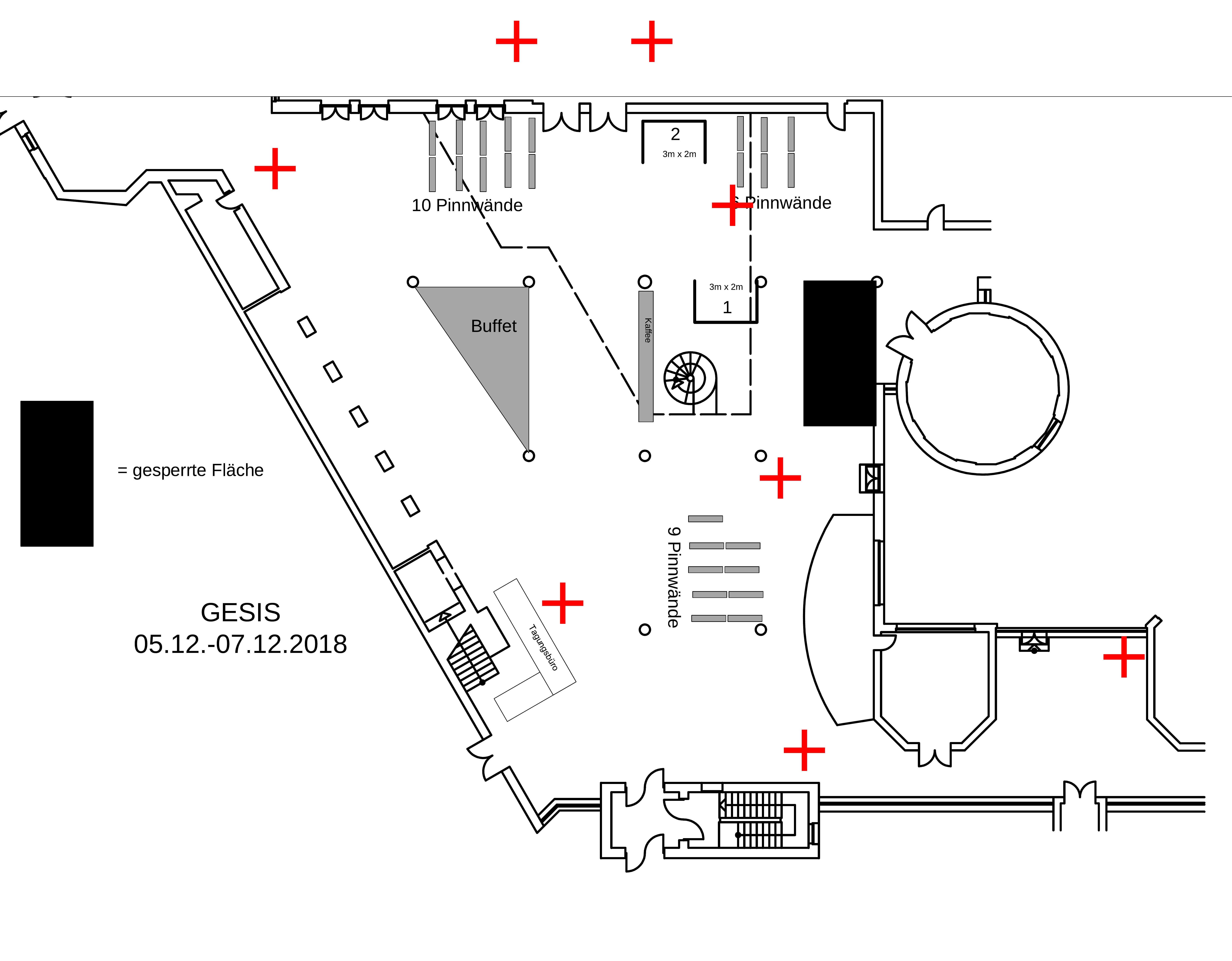}
    \caption{\textbf{Antenna locations for the ECSS18 conference.} ECSS18 was organised in the Maternushaus in Cologne, Germany. Antenna locations are indicated by red crosses. The main room is not visible on the plan, and was covered by the two antennas at the top. Basic floor plan by Maternushaus.}
    \label{fig:plans_ECSS18}
\end{figure*}

\begin{figure*}[h]
    \centering
    \includegraphics[width=.48\columnwidth]{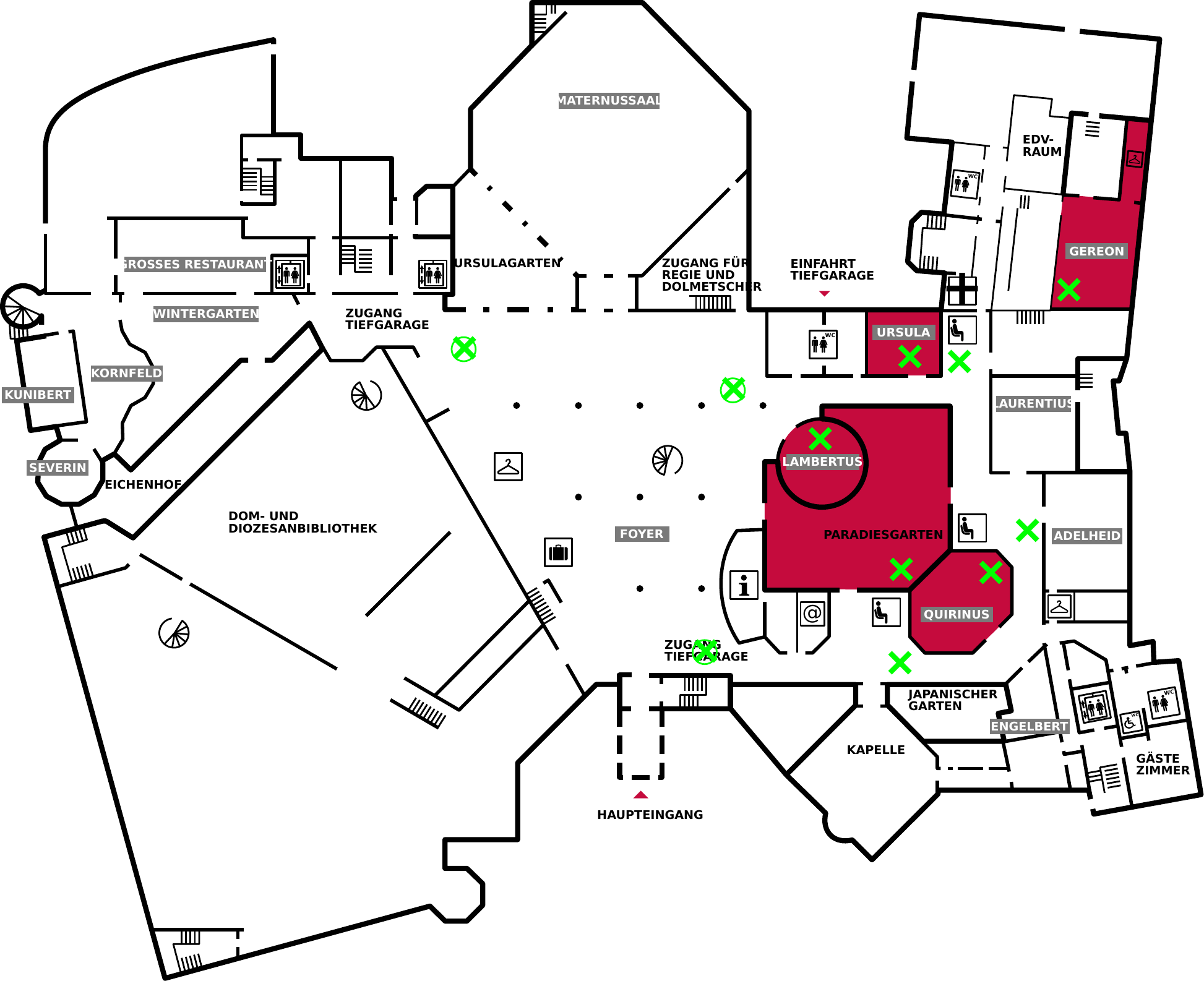}
    \includegraphics[width=.48\columnwidth]{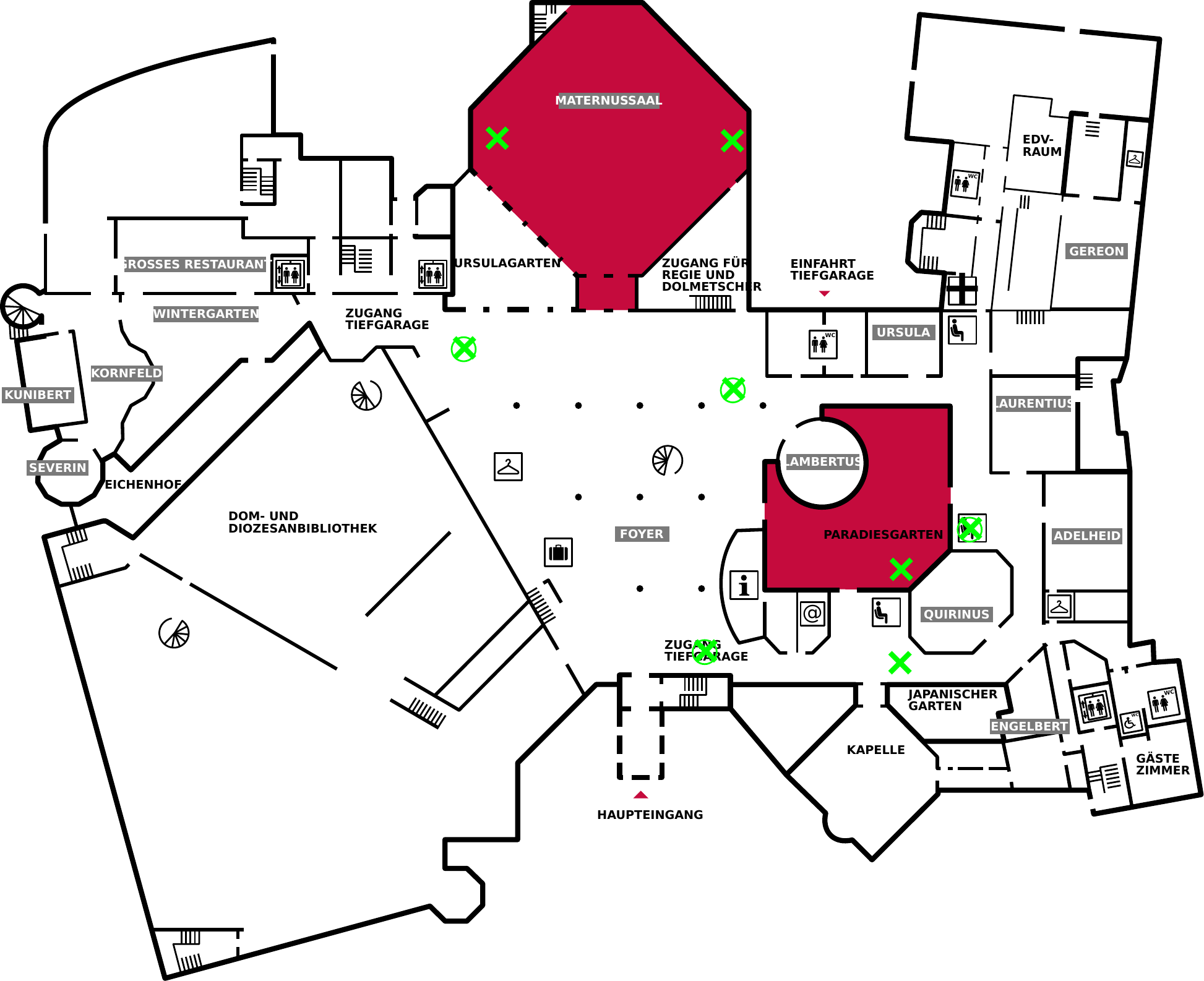}
    \\
    \includegraphics[width=.48\columnwidth]{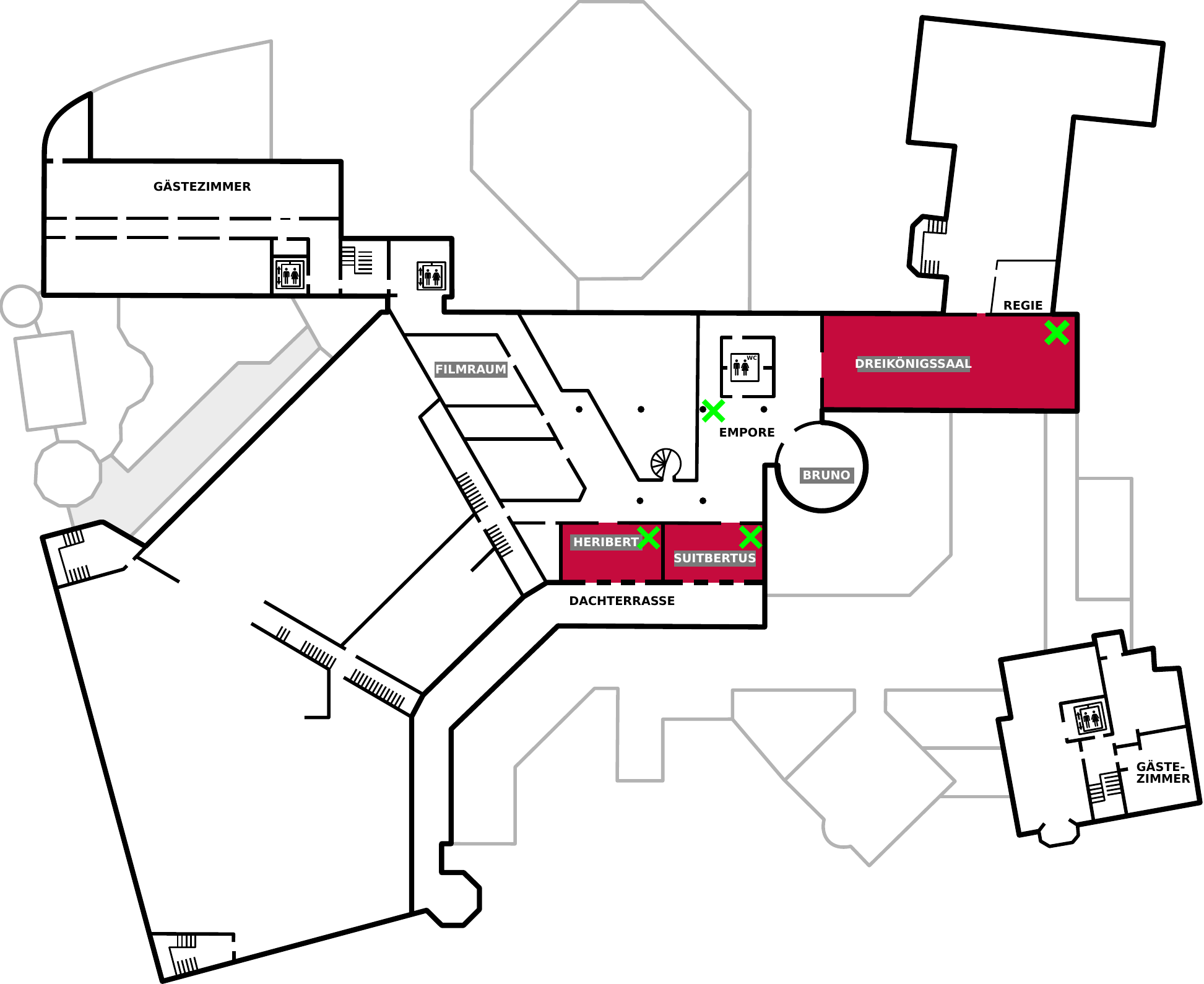}
    \includegraphics[width=.48\columnwidth]{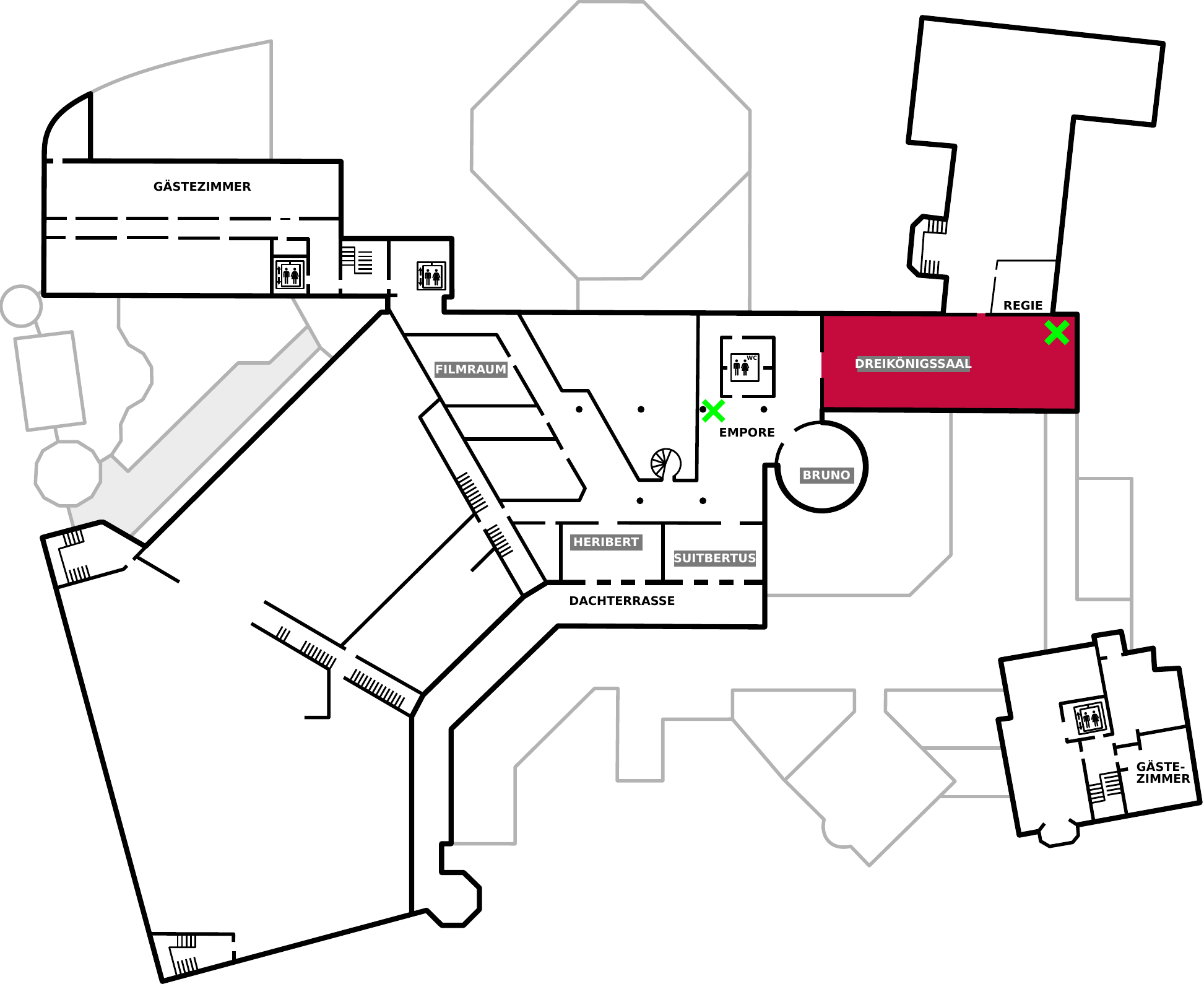}
    \caption{\textbf{Antenna locations for the ECIR19 conference.} ECIR19 was organised in the Maternushaus in Cologne, Germany. The venue consisted of two floors (top row: ground floor; bottom row: first floor). The conference had two parts: on Sunday (left column) workshops took place. Dark red areas indicate the rooms that were used. Basic floor plan by Maternushaus.}
    \label{fig:plans_ECIR19}
\end{figure*}

\clearpage
\section{Form of consent}\label{app:consent}

\begin{figure}[h!]
    \centering
    \includegraphics[width=.85\columnwidth]{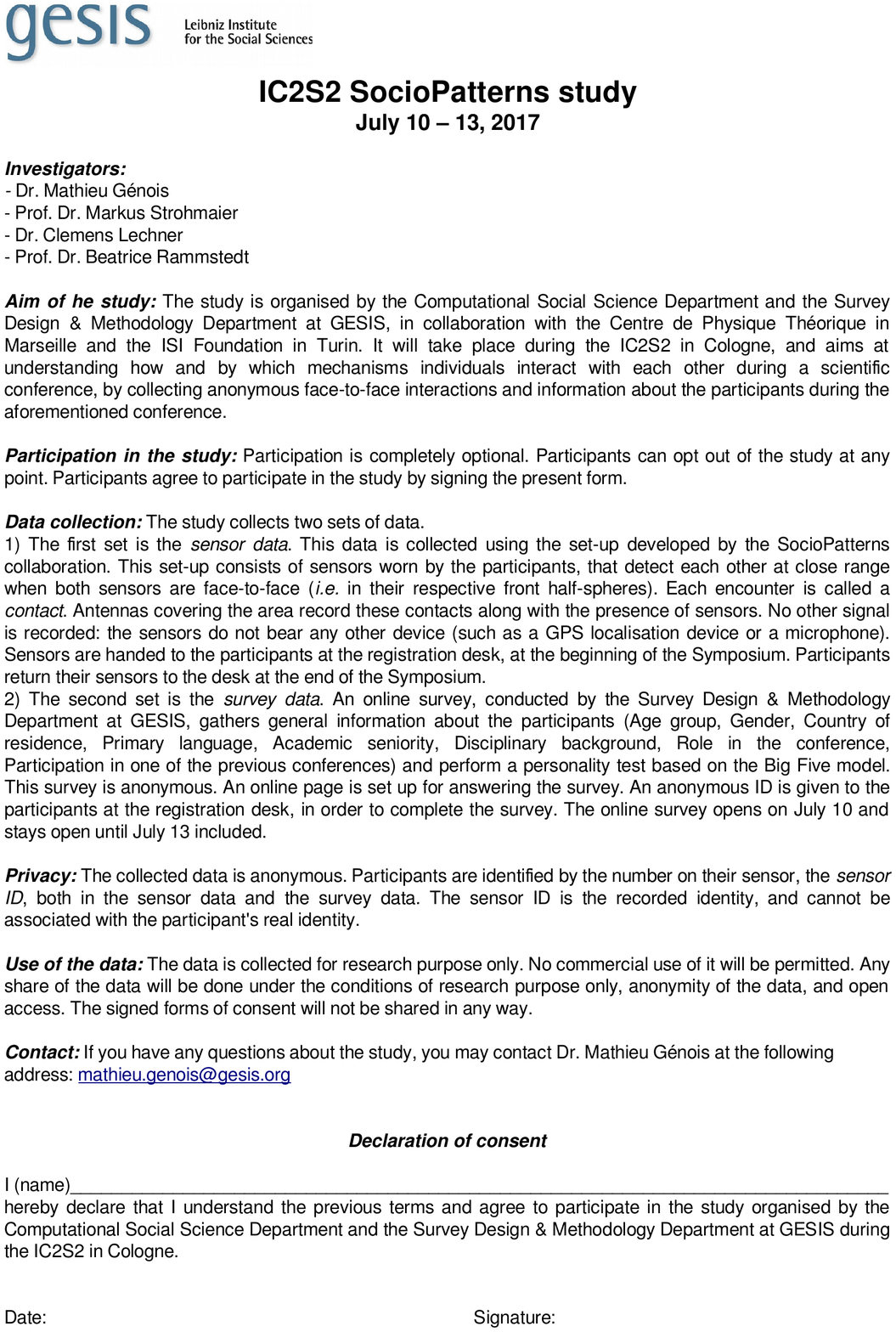}
    \label{fig:consent}
\end{figure}

\end{document}